\documentclass[a4paper,11pt]{article}

\usepackage{jcappub} 

\usepackage[T1]{fontenc} 
\usepackage{amsmath,amssymb}
\usepackage{bm}
\usepackage[normalem]{ulem}  
\graphicspath{{./fig/}}

\newcommand{\arXiv}[1]{\href{http://arxiv.org/abs/#1}}
\newcommand{\doi}[1]{\href{http://dx.doi.org/#1}}

\newcommand{\tr}{\mathrm{tr}\,}
\newcommand{\Tr}{\mathrm{Tr}\,}

\newcommand{\sgn}{\mathop{\mathrm{sgn}}}

\title{\boldmath Chiral gravitational waves from thermalized neutrinos in the early Universe}

\author[a]{Philipp Gubler,}
\author[b]{Naoki Yamamoto}
\author[c]{and Di-Lun Yang}

\affiliation[a]{Advanced Science Research Center, Japan Atomic Energy Agency, Tokai, Ibaraki 319-1195, Japan}
\affiliation[b]{Department of Physics, Keio University, Yokohama 223-8522, Japan}
\affiliation[c]{Institute of Physics, Academia Sinica, Taipei 11529, Taiwan}

\emailAdd{gubler@post.j-parc.jp}
\emailAdd{nyama@rk.phys.keio.ac.jp}
\emailAdd{dlyang@gate.sinica.edu.tw}

\abstract{
We investigate polarized gravitational waves generated by chiral fermions in the early Universe. In particular, we focus on the contribution from left-handed neutrinos in thermal equilibrium with finite temperature and chemical potential in the radiation dominated era. We compute the correlation functions of gravitational fields pertinent to the Stokes parameter $V$ characterizing the circular polarization of gravitational waves in the Minkowski and expanding spacetime backgrounds. In the expanding universe, we find that the thermalized neutrinos induce a non-vanishing $V$ linear to the neutrino degeneracy parameter and wavenumber of gravitational waves in the long wavelength region. 
While the magnitude of the gravitational waves generated by thermal neutrinos is too small to be detectable by current and planned third generation gravitational wave detectors, their observations by future generation detectors for ultra-high frequency regimes could provide information on the neutrino degeneracy parameter in the early Universe.
}

\begin{document}
\maketitle
\flushbottom

\section{Introduction}
\label{sec:intro}
With the experimental observation \cite{LIGOScientific:2016aoc} of gravitational waves in the ground based detectors LIGO \cite{LIGOScientific:2014pky} and Virgo \cite{VIRGO:2014yos}, 
while KAGRA \cite{KAGRA:2020cvd} is starting its operation, we are now in the era of gravitational wave astronomy. The gravitational waves observed so far, provide information about violent astronomical events such as black hole mergers or binary inspiraling neutron stars, but have also been used for constraining the energy density of 
the stochastic gravitational wave background \cite{LIGOScientific:2019vic}. 
With increasing precision of the existing facilities and future ground and space based detectors, it will furthermore become possible to measure the parity-odd components 
of gravitational waves \cite{Seto:2008sr,Gluscevic:2010vv,Hayama:2016kmv,Domcke:2019zls} (see also ref.\,\cite{Martinovic:2021hzy} for a recent analysis of the already available data). 
In the context of the early Universe, mechanisms that generate such parity violating gravitational waves have been discussed for instance in relation to pseudoscalar (or axion) inflation models with the gravitational topological coupling $\varphi R_{\mu \nu \alpha \beta} \tilde R^{\mu \nu \alpha \beta}$ \cite{Lue:1998mq,Choi:1999zy,Alexander:2004wk,Lyth:2005jf,Satoh:2007gn} or gauge topological coupling $\varphi F_{\mu \nu} \tilde F^{\mu \nu}$ \cite{Sorbo:2011rz,Cook:2011hg,Okano:2020uyr} and its non-Abelian version \cite{Adshead:2013qp,Maleknejad:2014wsa,Obata:2016tmo}, the existence of chiral fermions \cite{Anber:2016yqr,Barrie:2017mmr,Kamada:2021kxi}, the helical magnetic field \cite{Anand:2018mgf} due to the chiral plasma instability \cite{Joyce:1997uy,Akamatsu:2013pjd}, and helical turbulence \cite{Kahniashvili:2020jgm}.  

We will in this paper focus on (nearly) massless left-handed neutrinos in the early Universe that were thermalized after the inflation period.%
\footnote{Note that thermalized matter can 
emit gravitational waves due to the thermal fluctuations of energy-momentum tensors; see, e.g., ref.~\cite{Weinberg:1972kfs}. 
This is analogous to the generation of thermal photons 
from a neutral plasma \cite{Ghiglieri:2015nfa,Ghiglieri:2020mhm}.} 
Because these neutrinos have fixed left-handed chirality and violate parity within the Standard Model, they generate chiral gravitational waves with an imbalance between the two possible circular polarizations, which can be characterized by the Stokes parameter $V$. We especially consider the role of thermal effects, and demonstrate that in the regime where both the neutrino chemical potential $\mu$ and the momentum $k$ of the gravitational wave are much smaller than the temperature $T$ of the thermalized neutrino gas ($\mu/T$, $k/T \ll 1$), $V$ depends linearly on both $\mu/T$ and $k/T$; see eq.~(\ref{main}) for our main result. It is also straightforward to extend the results of the present paper to other types of chiral fermions or Dirac fermions with finite chiral chemical potential in thermal equilibrium. A deeper understanding of these chiral gravitational waves would also provide a baseline to study the possible parity-violating effects beyond the Standard Model.

This paper is organized as follows.
In section~\ref{sec:Stokes_para}, we briefly review the Stokes parameters for electromagnetic and gravitational waves. In section~\ref{sec:production}, we introduce a general framework for the production of gravitational waves from fermionic sources and obtain the correlations of gravitational fluctuations from chiral fermions in thermal equilibrium. In section~\ref{sec:GW_from_ neutrinos}, we adopt the developed formalism to investigate the birefringence of gravitational waves coming from thermalized left-handed neutrinos in both the Minkowski and expanding universes, where the Stokes parameters are evaluated and the relation to experimental measurements is discussed. Finally, a brief summary and outlook are presented in section~\ref{sec:summary}. 
The details of calculations are shown thereafter in multiple appendices. 

Throughout the paper, we use the natural units $\hbar = c = 1$ and mostly minus metric $\eta_{\mu \nu} = {\rm diag}(1, -1, -1, -1)$.
We also employ the shorthand notations $A_{(\mu}B_{\nu)}\equiv (A_{\mu}B_{\nu}+A_{\nu}B_{\mu})/2$ and $\int_{\bm p}\equiv \int\frac{{\rm d}^3p}{(2\pi)^3}$.

\section{The Stokes parameters for electromagnetic and gravitational waves}
\label{sec:Stokes_para}
We here briefly review the Stokes parameters, which were initially proposed for characterizing the polarization of 
electromagnetic waves \cite{Stokes:1852,Chandrasekhar:1950}, but can also be used to describe the polarization of gravitational waves \cite{Breuer:1972xs}. 

First, let us for illustration consider an electromagnetic wave moving in the direction of the z-axis, 
partially following ref.\,\cite{Piattella:2018hvi}. 
It suffices to consider only the properties of the electric field $\bm{E}$, as the magnetic field will be simply pointing to the direction perpendicular to 
the direction of the wave and $\bm{E}$. The electric field can be generally parametrized as 
\begin{align}
\bm{E} = \begin{pmatrix}
E_x \\
E_y \\
0
\end{pmatrix} 
= \begin{pmatrix}
A_x \cos(z - t) \\
A_y \cos(z - t + \beta) \\
0
\end{pmatrix} 
\end{align}
Eliminating the time variable, $E_x$ and $E_y$ can be shown to satisfy the equation 
\begin{align}
\frac{E_x^2}{A_x^2} + \frac{E_y^2}{A_y^2} - 2 \frac{E_x E_y}{A_x A_y} \cos \beta = \sin^2 \beta, 
\end{align}
which corresponds to a rotated ellipse in the $E_x$-$E_y$ plane. Defining $A_x = A \cos \theta$ and $A_y = A \sin \theta$, 
the rotation angle $\alpha$ of the ellipse is determined from
\begin{align}
\tan 2 \alpha = \tan 2 \theta \cos \beta, 
\end{align}
and the semi-major and semi-minor axes $a$ and $b$ are given as
\begin{align}
a^2 &= \frac{A^2}{2} \Bigl(1 + \sqrt{1 - \sin^2 2\theta \sin^2 \beta} \Bigr), \\
b^2 &= \frac{A^2}{2} \Bigl(1 - \sqrt{1 - \sin^2 2\theta \sin^2 \beta} \Bigr).
\end{align}
For $\beta = 0$, the wave is linearly polarized, while it is circularly polarized for $\beta = \pi/2$ and $\theta = \pi/4$. 

The nature of the polarization can be expressed using the Stokes parameters $I$, $Q$, $U$ and $V$, which are defined 
as 
\begin{align}
I &\equiv a^2 + b^2 = A^2, \label{eq:I_def} \\
Q &\equiv (a^2 - b^2) \cos 2\alpha = A^2 \cos 2 \theta, \\
U &\equiv (a^2 - b^2) \sin 2\alpha = A^2 \sin 2\theta \cos \beta, \\
V &\equiv 2 abh = A^2 h  \sin 2\theta \sin \beta, \label{eq:V_def}
\end{align}
where $h = \pm 1$ parametrizes the direction of rotation for the circularly polarized component of the wave. 
It is clear from the above definitions that $I$ gives the overall magnitude of the wave, $V$ the degree and direction of 
its circular polarization, and $Q$ and $U$ the degree and direction of the linear polarization of the wave. 
The four Stokes parameters are not independent, but are related to each other 
via 
\begin{align}
Q^2 + U^2 + V^2 = I^2. 
\label{eq:constraint}
\end{align}
Using the complex representation of an electric field, 
\begin{align}
\bm{E} = \begin{pmatrix}
E_x \\
E_y \\
0
\end{pmatrix} {\rm e}^{{\rm i}(z - t)},  
\end{align}
where $E_x$ and $E_y$ are complex and contain relative phases between the $x$ and $y$ component of the field, 
the Stokes parameters can be expressed as  
\begin{align}
I &\equiv |E_x|^2 + |E_y|^2, \\
Q &\equiv |E_x|^2 - |E_y|^2, \\
U &\equiv 2 \mathrm{Re}(E_x E_y^{\ast}), \\
V &\equiv -2\mathrm{Im}(E_x E_y^{\ast}), 
\end{align}
which can readily be confirmed to satisfy eq.\,(\ref{eq:constraint}) and to be consistent with the definitions of eqs.\,(\ref{eq:I_def})--(\ref{eq:V_def}) 
if one considers $\beta$ to be the relative phase of $E_x$ and $E_y$ and $\theta$ an angle determining the relative size of 
$|E_x|$ and $|E_y|$. Note that for a non-purely polarized wave, the above definitions should be regarded as expectation values over time of different 
components of $E_x$, $E_y$ and their complex conjugates. This can be cast in a concise form using the matrix
\begin{align}
\langle E_i E_j^{\ast} \rangle = \frac{1}{2} \begin{pmatrix}
I + Q & U - {\rm i}V \\
U + {\rm i}V & I - Q
\end{pmatrix},  
\label{eq:Stokes}
\end{align}
with $i$, $j = x$, $y$. 

In analogy to eq.\,(\ref{eq:Stokes}), the Stokes parameters of gravitational waves can be defined as \cite{Seto:2008sr}
\begin{align}
\langle h_i(\bm{k}) h_j^{\ast}(\bm{k}') \rangle = \frac{1}{2} \delta(\bm{k} - \bm{k}') \begin{pmatrix}
I(\bm{k}) + Q(\bm{k}) & U(\bm{k}) - {\rm i}V(\bm{k}) \\
U(\bm{k}) + {\rm i}V(\bm{k}) & I(\bm{k}) - Q(\bm{k})
\end{pmatrix},  
\label{eq:Stokes_gw}
\end{align}
where the subscripts $i$ and $j$ stand for the $+$ and $\times$ polarization modes of the 
gravitational wave in the transverse-traceless gauge. 
The $+$ and $\times$ components can be rewritten using right- and left-handed modes as 
\begin{align}
h_{\rm R} &= \frac{1}{\sqrt{2}} (h_{+} - {\rm i}h_{\times}), \\
h_{\rm L} &= \frac{1}{\sqrt{2}} (h_{+} + {\rm i}h_{\times}), 
\label{eq:circular_pol}
\end{align}
which modifies the above matrix to 
\begin{align}
\langle h_{\lambda}(\bm{k}) h_{\lambda'}^{\ast}(\bm{k}') \rangle = \frac{1}{2} \delta(\bm{k} - \bm{k}') \begin{pmatrix}
I(\bm{k}) + V(\bm{k}) & Q(\bm{k}) - {\rm i}U(\bm{k}) \\
Q(\bm{k}) + {\rm i}U(\bm{k}) & I(\bm{k}) - V(\bm{k})
\end{pmatrix},  
\end{align}
with $\lambda$, $\lambda' = {\rm R}$, ${\rm L}$. 
In particular, the $V$ parameter is explicitly given as the circular polarization of the gravitational wave:
\begin{eqnarray}
\label{eq:Stokes_circular}
\langle h_{\rm R}(\bm k,t)h_{\rm R}^{\ast}(\bm k',t)\rangle - \langle h_{\rm L}(\bm k,t)h_{\rm L}^{\ast}(\bm k',t)\rangle &=& \delta(\bm k - \bm k') V(\bm{k}).
\end{eqnarray}

Starting from the next section, we will for convenience of notation employ $\lambda$, $\lambda' = \pm1$ instead of 
$\lambda$, $\lambda' = {\rm R}$, ${\rm L}$. Especially, $\lambda$, $\lambda' =  +1$ should not be confused with the $+$ polarization mentioned above.
As we are mostly interested in parity violation components of the gravitational wave spectrum, our focus will be laid on the Stokes parameter $V$, which parametrizes the difference between the magnitudes of the right- and left-handed spectrum.

\section{Production of gravitational waves from thermal chiral fermions}
\label{sec:production}
In this section, we will demonstrate how to derive the two-point functions of gravitational fields induced by chiral fermions in the helicity basis. We will first review the general framework in subsection\,\ref{sec:general} and the application to the vacuum polarization in subsection\,\ref{sec:pol_vac}. Then, we generalize the results to compute thermal corrections in subsection\,\ref{sec:pol_thermal}.

\subsection{General framework}
\label{sec:general}
In setting up the general framework to study gravitational waves of differing helicities emitted from left-handed fermionic sources, 
we partly follow ref.~\cite{Anber:2016yqr}, but generalize it to take into account effects of finite temperature and 
chemical potential. The key objects that we wish to calculate are helical mode correlators of gravitational
perturbations generated by left-handed neutrinos. 

We start with the equation of motion (EOM) for gravitational perturbations $h_{\mu\nu}$ in the Friedmann-Lema\^{i}tre-Robertson-Walker (FLRW) background in the 
conformal coordinates, ${\rm d}s^2=a(\tau)^2({\rm d}\tau^2-{\rm d}\bm x^2)$. Here, $\tau$ is the conformal time, which is related to the cosmic time $t$ via ${\rm d} t = a {\rm d} \tau$, and $a(\tau)$ is the scale factor. (Later, we will also consider the Minkowski universe, which corresponds to setting $\tau=t$ and $a(\tau)=1$.)
We adopt the transverse-traceless (TT) gauge such that $h_{00}=h_{0i}=h^i_i=\partial_{j}h^j_i=0$, where $i,j$ denote the spatial coordinates. The EOM now reads%
\footnote{Here we neglect the damping term due to free-streaming neutrinos \cite{Weinberg:2003ur}, which may reduce the correlation of gravitational waves by several $10\%$ while this suppression does not change the order-of-magnitude estimate in our study.} 
\begin{eqnarray}\label{EOM_for_h}
\big(\partial_{\tau}^2+2\mathcal{H}\partial_{\tau}+k^2\big) h_{\mu\nu}({\bm k},\tau)=\frac{2}{m_{\rm P}^2}T_{\mu\nu}({\bm k},\tau),
\end{eqnarray}
where ${\cal H} \equiv ({\rm d}a/{\rm d}\tau)/a$ is the conformal Hubble parameter, $T_{\mu\nu}$ is the physical energy-momentum tensor (whose explicit form will be specified below), $m_{\rm P}=1/\sqrt{8\pi G}$ is the reduced Planck mass, and $k = |{\bm k}|$, $\bm k$ being the three momentum. 

In the TT gauge, $h_{\mu\nu}$ can be decomposed into circular helicity modes $h_{\pm}$ as
\begin{eqnarray}\label{h_decomp}
h^{\mu\nu}({\bm k},\tau)=\sum_{\lambda=\pm}\epsilon_{\lambda}^{\mu}({\bm k})\epsilon_{\lambda}^{\nu}({\bm k}) h_{\lambda}({\bm k},\tau),
\end{eqnarray}         
where $\epsilon^{\mu}_{\pm}$ denote the polarization four vectors with $\pm$ helicity satisfying
\begin{eqnarray}\label{pol_rel}
\epsilon_{\lambda}^{\mu~*} (\bm k) &=& \epsilon_{-\lambda}^{\mu}(\bm k)\,, \quad
\epsilon_{\lambda}^{\mu}(-\bm k) = - \epsilon_{-\lambda}^{\mu}(\bm k) \,,\quad
k \cdot \epsilon_{\lambda}(\bm k) = 0\,,\quad \nonumber \\
\epsilon_{\lambda}(\bm k) \cdot \epsilon_{\lambda'}^* (\bm k) &=& -\delta_{\lambda, \lambda'}, 
\quad \epsilon_{\lambda}(\bm k) \cdot \epsilon_{\lambda'} (\bm k) =-\delta_{\lambda,- \lambda'}.
\end{eqnarray}
Given the decomposition in eq.~(\ref{h_decomp}), the EOM for $h_{\lambda}$ yields
\begin{eqnarray}\label{EOM_ph}
\big(\partial_{\tau}^2+2\mathcal{H}\partial_{\tau}+k^2\big) h_{\lambda}({\bm k},\tau)=\frac{2}{m_{\rm P}^2}\epsilon^{\alpha\beta}_{-\lambda}(\bm k)\Pi^{\mu\nu}_{\quad,\alpha\beta}(\bm k)T_{\mu\nu}(\bm k,\tau),
\end{eqnarray}
where
\begin{eqnarray}
\epsilon^{\alpha\beta}_{\lambda}(\bm k) &\equiv& \epsilon_{\lambda}^{\alpha}(\bm k)\epsilon_{\lambda}^{\beta}(\bm k),
\quad
\Pi_{\mu\nu,\alpha\beta} (\bm k) \equiv {\cal P}_{\mu\alpha} (\bm k) {\cal P}_{\nu\beta} (\bm k) - \frac{1}{2}{\cal P}_{\mu\nu} (\bm k) {\cal P}_{\alpha\beta} (\bm k),
\quad \nonumber \\
{\cal P}_{\mu\nu}(\bm k) &\equiv& \delta_{\mu\nu} - \frac{k_\mu k_\nu}{k^2}.
\end{eqnarray}

The solution for $h_{\lambda}(\bm k, \tau)$ from the EOM of eq.~(\ref{EOM_ph}) can be written as
\begin{eqnarray}
h_{\lambda}(\bm k,\tau)=h_{\lambda}^{\text{hom}}(\bm k,\tau)-\frac{2}{m_{\rm P}^2}\epsilon^{\mu\nu}_{-\lambda}(\bm k)\int {\rm d}\tau'G_{k}(\tau,\tau')T_{\mu\nu}(\bm k,\tau'),
\end{eqnarray}
where  $\epsilon^{\alpha\beta}_{\lambda}(\bm k)\Pi^{\mu\nu}_{\quad,\alpha\beta}(\bm k)=\epsilon^{\mu\nu}_{\lambda}(\bm k)$ is used, $h_{\lambda}^{\text{hom}}(\bm k,t)$ is the homogeneous solution and $G_{k}(\tau,\tau')$ is the retarded Green's function satisfying 
\begin{eqnarray}\label{EOM_G}
\big(\partial_{\tau}^2+2\mathcal{H}\partial_{\tau}+k^2\big)G_{k}(\tau,\tau')=\delta(\tau-\tau').
\label{eq:GF}
\end{eqnarray} 
As $h_{\lambda}^{\text{hom}}(\bm k,\tau)$ is independent of $\lambda$ \cite{Anber:2016yqr}, we can focus on just the inhomogeneous solution. 
Accordingly, the correlation function of the helical modes of the gravitational waves reads
\begin{eqnarray}\label{h_correlation}
\langle h_{\lambda}(\bm k,\tau)h_{\lambda'}(\bm k',\tau)\rangle=\frac{4}{m_{\rm P}^4}\int {\rm d}\tau'{\rm d}\tau''G_{k}(\tau,\tau')G_{k'}(\tau,\tau'')M_{\lambda\lambda'}(\bm k,\bm k';\tau',\tau''),
\end{eqnarray}
where
\begin{eqnarray}\label{Mlambdal_def}
M_{\lambda\lambda'}(\bm k,\bm k';\tau',\tau'')=\epsilon^{\mu\nu}_{-\lambda}(\bm k)\epsilon^{\alpha\beta}_{-\lambda'}(\bm k')\langle T_{\mu\nu}(\bm k,\tau')T_{\alpha\beta}(\bm k',\tau'')\rangle.
\end{eqnarray}

For the symmetric energy-momentum tensor, we consider the contribution of left-handed fermions (neutrinos),
\begin{eqnarray}
T_{\mu\nu}(x)=\frac{1}{a^2}
\left( \frac{\rm i}{2}\big[\psi^{\dagger}(x)\bar{\sigma}_{(\mu}\partial_{\nu)}\psi(x)-\partial_{(\mu}\psi^{\dagger}(x)\bar{\sigma}_{\nu)}\psi(x)\big] 
- {\rm i} \eta_{\mu \nu} \psi^{\dagger}(x)\bar{\sigma}^{\lambda}\partial_{\lambda}\psi(x)  \right)\,,
\end{eqnarray}
where $\sigma^{\mu} = (1, {\bm \sigma})$, $\bar \sigma^{\mu} = (1, -{\bm \sigma})$, $\sigma^i$ being Pauli matrices, $\psi$ the rescaled fermionic field, and $x^{\mu}=(\tau,\bm x)$. Note that the term proportional to $\eta_{\mu \nu}$ will be irrelevant when we consider the transverse part of $T_{\mu \nu}$ in eq.~(\ref{EOM_ph}) in momentum space. The mode expansion of the $\psi$ field takes the form 
\begin{eqnarray}\label{mode_exp}
\psi(x)=
\int_{\bm p} \left[\xi_{-}(\bm p)u(\tau,\bm p)a_{\bm p}+\xi_{-}(-\bm p)v(\tau,-\bm p)b^{\dagger}_{-\bm p}\right]{\rm e}^{{\rm i}\bm p\cdot \bm x},
\end{eqnarray}
where $\xi_-(\bm p)$ is the negative eigenstate of the helicity operator such that 
\begin{eqnarray}\label{xi_rel}
{\bm \sigma} \cdot {\bm p} \xi_{-}(\bm p) &=& -p\xi_-(\bm p),\quad \xi_{-}^{\dagger}(\bm p)\xi_{-}(\bm p)={\bm 1}, \nonumber \\
 \xi_{-}(\bm p)\xi_{-}^{\dagger}(\bm p) &=& \frac{1}{2}\sigma\cdot n(\bm p),
\quad n^{\mu}=(1,{\bm p}/p),
\end{eqnarray}
and $a_{\bm p}$ and $b^{\dagger}_{\bm p}$ denote the annihilation operator for fermions and creation operator for anti-fermions, respectively. They follow the fermionic commutation relations
\begin{eqnarray}
\{a_{\bm q},a^{\dagger}_{\bm p}\}=(2\pi)^3\delta^3({\bm q}-{\bm p}),\quad \{b_{\bm q},b^{\dagger}_{\bm p}\}=(2\pi)^3\delta^3({\bm q}-{\bm p}),\quad 
\{a_{\bm q},a_{\bm p}\}=\{b_{\bm q},b_{\bm p}\}=0.
\end{eqnarray}
Using the mode expansion in eq.~(\ref{mode_exp}), we find
\begin{eqnarray}\label{Tmunu_k}
T_{\mu\nu}(\bm k,\tau)=\frac{1}{2a(\tau)^{2}}\int {\rm d}^3x \int_{\bm p}\int_{\bm p'} L^{\dagger}(\bm p', \tau )\bar{\sigma}_{(\mu}\big(p_{\nu)}+p'_{\nu)}\big)L(\bm p,\tau){\rm e}^{{\rm i}({\bm p- \bm p'- \bm k})\cdot\bm x},
\end{eqnarray}
where  
\begin{eqnarray}
L(\bm p,\tau)\equiv \xi_{-}(\bm p)u(\tau,\bm p)a_{\bm p}+\xi_{-}(-\bm p)v(\tau,-\bm p)b^{\dagger}_{-\bm p}.
\end{eqnarray} 
As shown in ref.~\cite{Anber:2016yqr}, by inserting eq.~(\ref{Tmunu_k}) into eq.~(\ref{h_correlation}), one obtains
\begin{eqnarray}\nonumber\label{Mlambda}
M_{\lambda\lambda'}(\bm k,\bm k';\tau',\tau'')&=&\frac{1}{4a(\tau')^2a(\tau'')^2}
\int_{\bm p}\int_{\bm p'}
\Big\langle L^{\dagger}({\bm p - \bm k},\tau')\epsilon^{\mu\nu}_{-\lambda}(\bm k)
\\
&&\times 
\big[\bar{\sigma}_{\mu}p_{\nu}+\bar{\sigma}_{\mu}(p-k)_{\nu}\big]
L(\bm p, \tau')L^{\dagger}({\bm p' - \bm k'},\tau'')\epsilon^{\rho\sigma}_{-\lambda'}(\bm k') \nonumber
\\
&&\times 
\big[\bar{\sigma}_{\rho}p'_{\sigma}+\bar{\sigma}_{\rho}(p'-k')_{\sigma}\big]L(\bm p',\tau'')\Big\rangle.
\end{eqnarray}

\subsection{Polarization in vacuum}\label{sec:pol_vac}
When considering just the vacuum expectation value, the only non-vanishing combination of the creation and annihilation operators 
(besides the trivial $\bm k=\bm k'= \bm 0$ case, which we ignore here) is
\begin{eqnarray}\label{vac_comb}
\langle 0|b_{\bm k - \bm p}a_{\bm p}a^{\dagger}_{\bm p' - \bm k'}b^{\dagger}_{-\bm p'}|0\rangle 
&=&(2\pi)^6\delta^3({\bm p- \bm p'+ \bm k'})\delta^3({\bm p'- \bm p + \bm k}) \nonumber \\
&=&(2\pi)^6\delta^3({\bm k + \bm k'})\delta^3({\bm p'- \bm p + \bm k}).
\end{eqnarray}  
We hence have 
\begin{eqnarray}\nonumber\label{M_in_beta}
&&M_{\lambda\lambda'}(\bm k,\bm k';\tau',\tau'')
\\
&&=(2\pi)^3\frac{\delta^3({\bm k + \bm k'})}{a(\tau')^2a(\tau'')^2}\int_{\bm p}v^{\dagger}({\bm p - \bm k},\tau')u(\bm p,\tau')u^{\dagger}({\bm p' - \bm k'},\tau'')v(\bm p',\tau'')\beta_{\lambda\lambda'}({\bm k, \bm p})\,,
\end{eqnarray}
where 
\begin{eqnarray}
&&\beta_{\lambda\lambda'}({\bm k, \bm p}) \nonumber \\
&&=\big(\epsilon_{-\lambda}(\bm k)\cdot p\big)\big(\epsilon_{\lambda'}(\bm k)\cdot p\big) \epsilon^{\mu}_{-\lambda}(\bm k)\epsilon^{\rho}_{\lambda'}(\bm k)
\xi^{\dagger}_{-}({\bm k - \bm p})\bar{\sigma}_{\mu}\xi_{-}(\bm p)
\xi^{\dagger}_{-}({\bm p})\bar{\sigma}_{\rho}\xi_{-}({\bm k - \bm p}).
\end{eqnarray}
Making use of eq.~(\ref{xi_rel}), $\beta_{\lambda\lambda'}({\bm k, \bm p})$ can be further rewritten as
\begin{eqnarray}\nonumber\label{beta_eq1}
\beta_{\lambda\lambda'}({\bm k, \bm p})
&=&\frac{1}{4}\big(\epsilon_{-\lambda}(\bm k)\cdot p\big)\big(\epsilon_{\lambda'}(\bm k)\cdot p\big)\epsilon^{\mu}_{-\lambda}(\bm k)\epsilon^{\rho}_{\lambda'}(\bm k)n^{\alpha}({\bm k - \bm p})n^{\beta}(\bm p)
\tr\big(\sigma_{\alpha}\bar{\sigma}_{\mu}\sigma_{\beta}\bar{\sigma}_{\rho}\big)
\\\nonumber
&=&-\frac{1}{2}\big(\epsilon_{-\lambda}(\bm k)\cdot p\big)\big(\epsilon_{\lambda'}(\bm k)\cdot p\big)
\Bigg[\frac{2\big(\epsilon_{-\lambda}(\bm k)\cdot p\big)\big(\epsilon_{\lambda'}(\bm k)\cdot p\big)}{p|{\bm k - \bm p}|}
-\delta_{\lambda,\lambda'}\left(1+\frac{p^2-{\bm k\cdot \bm p}}{p|{\bm k - \bm p}|}\right)
\\
&&+{\rm i}\epsilon_{\mu\beta\rho\alpha}
\epsilon^{\mu}_{-\lambda}(\bm k)\epsilon^{\rho}_{\lambda'}(\bm k)n^{\alpha}({\bm k - \bm p})n^{\beta}(\bm p)
\Bigg]
,
\end{eqnarray}
where we utilized 
\begin{eqnarray}
\tr\big(\sigma_{\alpha}\bar{\sigma}_{\mu}\sigma_{\beta}\bar{\sigma}_{\rho}\big)=2\big(\eta_{\mu\beta}\eta_{\rho\alpha}-\eta_{\mu\rho}\eta_{\beta\alpha}+\eta_{\mu\alpha}\eta_{\beta\rho}-{\rm i}\epsilon_{\mu\beta\rho\alpha}\big)
\end{eqnarray}
with $\epsilon_{0123}=-1$. 
The last (imaginary) term in eq.~(\ref{beta_eq1}) is critical to break parity. 

To integrate over $\bm p$, one has to explicitly fix the coordinate system. 
Following ref.~\cite{Anber:2016yqr}, we take $\bm k$ along $z$ (or equivalently $x^3$) direction and explicitly define the polarization vectors as
\begin{eqnarray}
\epsilon^{\mu}_{\lambda}(\bm k)=\frac{1}{\sqrt{2}}(0,\sgn(k^z),{\rm i}\lambda,0),
\end{eqnarray} 
where $\sgn(k^z)$ is introduced to satisfy the relation $\epsilon_{\lambda}^{\mu}(-\bm k) = - \epsilon_{-\lambda}^{\mu}(\bm k)$ in eq.~(\ref{pol_rel}). We will focus on $k^z>0$ such that $k^{\mu}=(k,0,0,k)$ without loss of generality. 
$\bm p$ is parametrized as $\bm p = p(\sin \theta \cos \phi, \sin \theta \sin \phi, \cos \phi)$. 
Now, $\beta_{\lambda\lambda'}({\bm k, \bm p})$ in spherical coordinates can be given as
\begin{eqnarray}\nonumber\label{beta_eq2}
\beta_{\lambda\lambda'}({\bm k, \bm p})=&&\frac{p^2}{4}\sin^2\theta\big(\cos^2\phi+\lambda\lambda'\sin^2\phi+{\rm i}(\lambda'-\lambda)\cos\phi\sin\phi\big)
\Bigg[\delta_{\lambda,\lambda'}-\frac{\lambda+\lambda'}{2}\cos\theta \nonumber \\
&& +\frac{1}{|{\bm k - \bm p}|}\bigg(\delta_{\lambda,\lambda'}
\big(p-k\cos\theta\big)+\frac{\lambda+\lambda'}{2}\big(k-p\cos\theta\big) \nonumber \\
&& -p\sin^2\theta\big(\cos^2\phi+\lambda\lambda'\sin^2\phi+{\rm i}(\lambda'-\lambda)\cos\phi\sin\phi\big)
\bigg)\Bigg],
\end{eqnarray}
which yields%
\footnote{Here, $\beta_{\lambda\lambda}(\bm k, \bm p)$ is different from the result found in ref.~\cite{Anber:2016yqr} with opposite signs in front of $\lambda$.}
\begin{align}
\label{beta_same_helicity}
\beta_{\lambda\lambda}({\bm k, \bm p})
&=\frac{p^2\sin^2\theta}{4}
\Bigg[1-\lambda\cos\theta
+(\cos\theta-\lambda)\frac{p\cos\theta-k}{|{\bm k - \bm p}|}
\Bigg]\,,
\\
\beta_{\lambda,-\lambda}({\bm k, \bm p})
&=-\frac{p^3\sin^4\theta\big(\cos 4\phi-{\rm i}\lambda\sin 4\phi\big)}{4|{\bm k - \bm p}|}\,,
\end{align}
where we used $\lambda^2=1$.
Note that $\beta_{\lambda,-\lambda}({\bm k, \bm p})$, which is the only term depending on $\phi$ in the integrand of eq.~(\ref{M_in_beta}), 
does not contribute to $M_{\lambda\lambda'}(\bm k,\bm k';\tau',\tau'')$ by symmetry. 
Now, inserting eq.~(\ref{beta_same_helicity}) into eq.~(\ref{M_in_beta}), the correlation function in eq.~(\ref{h_correlation}) becomes  
\begin{eqnarray}\label{h_correlation_vac}\nonumber
&& \langle h_{\lambda}(\bm k,\tau)h_{\lambda'}(\bm k',\tau)\rangle \\
&&=\delta_{\lambda,\lambda'}\frac{\delta^3({\bm k + \bm k'})}{m_{\rm P}^4}\int {\rm d}\tau'{\rm d}\tau''\frac{G_{k}(\tau,\tau')G_{k'}(\tau,\tau'')}{a(\tau')^2a(\tau'')^2}
\int {\rm d}^3p
v^{\dagger}({\bm p - \bm k},\tau')u(\bm p,\tau') \nonumber \\
&&\quad \times u^{\dagger}({\bm p' - \bm k'},\tau'')v(\bm p',\tau'')
p^2\sin^2\theta\Bigg[1-\lambda\cos\theta
+(\cos\theta-\lambda)\frac{p\cos\theta-k}{|{\bm k - \bm p}|}
\Bigg].
\end{eqnarray}

\subsection{Thermal corrections}\label{sec:pol_thermal}
So far, we have closely followed ref.~\cite{Anber:2016yqr} in our derivation of the correlation function. The purpose of this work, however, 
is to study not the effect of vacuum fluctuations of left-handed neutrinos, but to evaluate their thermal contributions to the generation of gravitational waves in the early Universe before decoupling. Despite our focus on neutrinos, the following approach can be applied to chiral fermions in general. 

We first introduce the thermal average, which is defined for a generic operator $\hat O$ as
\begin{eqnarray}
\langle\hat{O}\rangle &\equiv& \Tr\big(\hat{O}\hat{\rho}\big),\quad \hat{\rho}=\frac{1}{Z}{\rm e}^{-\beta\big[(E_q-\mu)\hat{n}_{\bm q}+(E_q+\mu)(\hat{\tilde{n}}_{\bm q}-1)\big]}\,, \nonumber \\
Z &\equiv& \Tr\left({\rm e}^{-\beta\left[(E_q-\mu)\hat{n}_{\bm q}+(E_q+\mu)(\hat{\tilde{n}}_{\bm q}-1)\right]}\right),
\label{eq:thermal_equib}
\end{eqnarray} 
where $\hat{n}_{\bm q}\equiv (a^{\dagger}_{\bm q}a_{\bm q})/{\cal V}$ and $\hat{\tilde{n}}_{\bm q}\equiv (b^{\dagger}_{\bm q}b_{\bm q})/{\cal V}$ with ${\cal V}\equiv (2\pi)^3\delta^3(\bm 0)$ 
denote the normalized density operators for neutrinos and anti-neutrinos and $\beta=1/T$.
For the case of the expanding universe that we will consider later, all dimensionful quantities, such as the physical temperature $T$ and chemical potential $\mu$ (as well as momenta) will be rescaled by the scale factor $a$ as $\bar T = a T$ and $\bar \mu = a \mu$, when taking the thermal average. As $T(\tau)\propto 1/a(\tau)$ and $\mu(\tau) \propto 1/a(\tau)$, $\bar T$ and $\bar \mu$ are constants as a function of $\tau$. Nonetheless, for notational simplicity, we will use ${\bm k}$ etc. to denote the rescaled momentum. The physical momentum depending on $\tau$ is given by ${\bm k}_{\rm phys}(\tau) = {\bm k}/a(\tau)$. The traces in eq.\,(\ref{eq:thermal_equib}) are taken for vacuum and one-particle (and anti-particle) states, 
while states with more than one particle are forbidden from the Pauli exclusion principle. 
Specifically, we define
\begin{eqnarray}
\Tr(\hat{O})=\langle 0|\hat{O}|0\rangle+\int_{\bm q}\langle 0|a_{\bm q}\hat{O}a_{\bm q}^{\dagger}|0\rangle+\int_{\bm q}\langle 0|b_{\bm q}\hat{O}b_{\bm q}^{\dagger}|0\rangle
+\frac{1}{\cal V} \int_{\bm q}\langle 0|a_{\bm q}b_{\bm q}\hat{O}b_{\bm q}^{\dagger}a_{\bm q}^{\dagger}|0\rangle
\label{eq:trace}
\end{eqnarray}
and $\int_{\bm q}{\cal V}\equiv 1$.
For this, we find
\begin{eqnarray}
Z=1+{\rm e}^{-\beta(E_q-\mu)}+{\rm e}^{\beta(E_q+\mu)}+{\rm e}^{-\beta(E_q-\mu)+\beta(E_q+\mu)}=\frac{1}{(1-f_{E_q-\mu})f_{E_q+\mu}}\,,
\end{eqnarray}
where $f_{q}\equiv 1/({\rm e}^{\beta q}+1)$. 
Furthermore, one can easily confirm that the expectation values of density operators give rise to the Fermi-Dirac distribution functions,
\begin{eqnarray}\nonumber
\langle \hat{n}_{\bm p}\rangle&=&\int_{\bm q}(2\pi)^3\delta^3({\bm q - \bm p})\frac{1}{Z}
{\rm e}^{-\beta (E_q-\mu)}\bigl[{\rm e}^{\beta(E_q+\mu)}+1 \bigr]
=\frac{1}{{\rm e}^{\beta(E_p-\mu)}+1}\,,
\\
\langle \hat{\tilde{n}}_{\bm p}\rangle&=&\int_{\bm q}(2\pi)^3\delta^3({\bm q - \bm p})\frac{1}{Z}
\bigl[{\rm e}^{-\beta(E_q-\mu)}+1 \bigr]
=\frac{1}{{\rm e}^{\beta(E_p+\mu)}+1}\,.
\end{eqnarray}

We now replace the vacuum expectation value of eq.~(\ref{Mlambda}) with the trace of eq.\,(\ref{eq:thermal_equib}), to incorporate the thermal excitations 
and the effects of the chemical potential, and look for combinations of operators, which have non-trivial expectation values. This will include the vacuum part that we have already computed in the previous subsection. 
As the details of this calculation are rather involved, we relegate them to appendix\,\ref{sec:app_thermal_part}, and here show only the final result, 
which can be divided into the vacuum and thermal parts, 
\begin{eqnarray}
\langle h_{\lambda}(\bm k,\tau)h_{\lambda'}(\bm k',\tau)\rangle\equiv \langle h_{\lambda}(\bm k,\tau)h_{\lambda'}(\bm k',\tau)\rangle_{\text{vac}}+\langle h_{\lambda}(\bm k,\tau)h_{\lambda'}(\bm k',\tau)\rangle_{\text{th}},
\end{eqnarray}   
where $\langle h_{\lambda}(\bm k,\tau)h_{\lambda'}(\bm k',\tau)\rangle_{\text{vac}}$ is given in eq.~(\ref{h_correlation_vac}), which generally 
suffers from an ultraviolet divergence and may have to be renormalized by subtracting proper counter terms or by introducing a cutoff \cite{Anber:2016yqr}. 
For the thermal part, after the rescaling of dimensionful quantities, we have
\begin{eqnarray}\nonumber\label{eq:hcor_th}
\langle h_{\lambda}(\bm k,\tau)h_{\lambda'}(\bm k',\tau)\rangle_{\text{th}} 
&=&\delta_{\lambda,\lambda'}\frac{\delta^3({\bm k + \bm k'})}{m_{\rm P}^4}\int {\rm d}\tau'{\rm d}\tau''\frac{G_{k}(\tau,\tau')G_{k'}(\tau,\tau'')}{a(\tau')^2a(\tau'')^2}\int {\rm d}^3p \ p^2 \nonumber \\
&& \times \Big[\Gamma_{+}(\bm p, \theta;\bm k,\lambda;\tau',\tau'')f_{|{\bm k - \bm p}|-{\bar \mu}}
+\Gamma_{-}(\bm p, \theta;\bm k,\lambda;\tau',\tau'')f_{p+\bar \mu} \nonumber \\
&&-\Gamma_{0}(\bm p, \theta;\bm k,\lambda;\tau',\tau'')\big(f_{|{\bm k - \bm p}|+\bar \mu}+f_{p- \bar \mu}\big)\Big],
\end{eqnarray}
where
\begin{align}
\label{eq:Gamma_pm}
\Gamma_{\pm}(\bm p, \theta;\bm k,\lambda;\tau',\tau'') 
&= {\rm e}^{{\rm i}(|{\bm k - \bm p}|-p)\tau'}{\rm e}^{{\rm i}(p-|{\bm k - \bm p}|)\tau''}
\tilde{\Gamma}_{\pm}(\bm p, \theta;\bm k,\lambda), \\
\label{eq:Gamma_0}
\Gamma_{0}(\bm p, \theta;\bm k,\lambda;\tau',\tau'') 
&={\rm e}^{-{\rm i}(|{\bm k - \bm p}|+p)\tau'}{\rm e}^{{\rm i}(p+|{\bm k - \bm p}|)\tau''}
\tilde{\Gamma}_{0}(\bm p, \theta;\bm k,\lambda),
\end{align}
with
\begin{align}
\label{eq:Gamma_bar_pm}
\tilde{\Gamma}_{\pm}(\bm p, \theta;\bm k,\lambda) &=
\sin^2\theta\Bigg[1\mp \lambda\cos\theta
-(\cos\theta \mp \lambda)\frac{p\cos\theta-k}{|{\bm k - \bm p}|}
\Bigg], \\
\label{eq:Gamma_bar_0}
\tilde{\Gamma}_{0}(\bm p, \theta;\bm k,\lambda) &=
\sin^2\theta\Bigg[1-\lambda\cos\theta
+(\cos\theta-\lambda)\frac{p\cos\theta-k}{|{\bm k - \bm p}|}
\Bigg].
\end{align}

\section{Gravitational birefringence from primordial neutrinos}\label{sec:GW_from_ neutrinos}
Utilizing the formalism introduced in the previous section, we next explicitly evaluate the Stokes parameter due to primordial neutrinos in thermal equilibrium. As a heuristic study, we first investigate the case of the Minkowski universe in subsection\,\ref{sec:flat}. Subsequently, we analyze the more practical case for an expanding universe in the radiation dominated period in subsection\,\ref{sec:expanding}.
\subsection{Minkowski universe}
\label{sec:flat}
In the Minkowski spacetime, where $\tau=t$, $a(\tau)=1$, and $\mathcal{H}(\tau)=0$, eq.~(\ref{EOM_G}) simply results in 
\begin{eqnarray}
G_k(t,t') &=& -\int \frac{{\rm d}k_0}{2\pi}{\rm e}^{-{\rm i}k_0(t-t')}\frac{1}{(k_0 + i \epsilon)^2-k^2} \nonumber \\
&=& \theta(t-t') \frac{1}{k} \sin k(t-t')\,.
\end{eqnarray}
Moreover, the fermionic wave functions reduce to the plane waves,
\begin{eqnarray}
u(t,p)={\rm e}^{-{\rm i}p\cdot x},\quad v(t,p)={\rm e}^{{\rm i}p\cdot x}.
\end{eqnarray}
One can then easily evaluate the integrations with respect to $t'$ and $t''$ in eq.~(\ref{h_correlation_vac}) and derive the correlation function in vacuum,
\begin{eqnarray}\label{h_correlation_vac_flat}\nonumber
	\langle h_{\lambda}(\bm k,t)h_{\lambda'}(\bm k',t)\rangle
	&=&\delta_{\lambda,\lambda'}\frac{\delta^3({\bm k + \bm k'})}{4m_{\rm P}^4}\int {\rm d}^3p\frac{\sin^2\theta}{\big(p-k\cos\theta+|{\bm k - \bm p}|\big)^2}
	\\
	&&\times\Bigg[1-\lambda\cos\theta
	+(\cos\theta-\lambda)\frac{p\cos\theta-k}{|{\bm k - \bm p}|}
	\Bigg]
	,
\end{eqnarray}
where $|{\bm k - \bm p}|=\sqrt{p^2-2pk\cos\theta+k^2}$. The correlation function is in fact independent of $t$ since now we work in the static background spacetime.
  
For the thermal part, by carrying out similar computations, eq.~(\ref{eq:hcor_th}) becomes
\begin{eqnarray}\nonumber\label{hcor_th}
	\langle h_{\lambda}(\bm k,t)h_{\lambda'}(\bm k',t)\rangle_{\text{th}}&=&
	\delta_{\lambda,\lambda'}\frac{\delta^3({\bm k + \bm k'})}{4m_{\rm P}^4}\int {\rm d}^3p\Big[\Pi_{+}(\bm p, \theta;\bm k,\lambda)f_{|{\bm k - \bm p}|-\mu}
	+\Pi_{-}(\bm p, \theta;\bm k,\lambda)f_{p+\mu}
	\\
	&&-\Pi_{0}(\bm p, \theta;\bm k,\lambda)\big(f_{|{\bm k - \bm p}|+\mu}+f_{p-\mu}\big)\Big]\,,
\end{eqnarray}
where
\begin{align}
	\Pi_{\pm}(\bm p, \theta;\bm k,\lambda)&=\frac{\sin^2\theta}{\big(p-k\cos\theta-|{\bm k - \bm p}|\big)^2}\left[1\mp\lambda\cos\theta
	-(\cos\theta\mp\lambda)\frac{p\cos\theta-k}{|{\bm k - \bm p}|}
	\right]\,, \\
	\Pi_{0}(\bm p, \theta;\bm k,\lambda)&=\frac{\sin^2\theta}{\big(p-k\cos\theta+|{\bm k - \bm p}|\big)^2}\left[1-\lambda\cos\theta
	+(\cos\theta-\lambda)\frac{p\cos\theta-k}{|{\bm k - \bm p}|}
	\right]\,.
\end{align} 
When considering the long-wavelength (small $k$) limit, the $p$ integral can be computed analytically. 
In this limit, one finds $\Pi_{\pm}(\bm p, \theta;\bm k,\lambda)$ to be the dominant terms. 
In particular, we find 
\begin{eqnarray}
	\Pi_{+}(\bm p, \theta;\bm k,\pm 1)\approx \left[\frac{k^4}{4p^2}\pm\frac{\cos^2(\theta/2)k^5}{2p^3}+\mathcal{O}\left(\frac{k^6}{p^4}\right)\right]^{-1}\approx \frac{4p^2}{k^4}\mp \frac{8p\cos^2(\theta/2)}{k^3}\, ,
\end{eqnarray} 
and can approximate $f_{|{\bm k - \bm p}|-\mu}\rightarrow f_{p-\mu}$.

Interestingly, when taking $\mu\gg T$ despite being an unrealistic condition in the early Universe, the Fermi-Dirac distribution function reduces to a unit-step function, which leads to $f_{p-\mu}\rightarrow \Theta(\mu-p)$ and $f_{p+\mu}\rightarrow \Theta(\mu+p)=0$ for $\mu>0$. In such a case, only $\Pi_{+}(\bm p, \theta;\bm k,\pm 1)$ contributes for $k \ll \mu$ and eq.~(\ref{hcor_th}) approximately becomes   
\begin{eqnarray}\nonumber\label{hcor_th_twolimits}
	\langle h_{\pm}(\bm k,t)h_{\pm}(\bm k',t)\rangle_{\text{th}}&\approx&\frac{2\pi\delta^3({\bm k + \bm k'})}{4m_{\rm P}^4}\int^{\infty}_{0} {\rm d}p \int^{\pi}_0 {\rm d}\theta p^2\sin\theta\left[\frac{4p^2}{k^4}\mp\frac{8p\cos^2(\theta/2)}{k^3}\right]\Theta(\mu-p)
	\\
	&=&\frac{\pi\delta^3({\bm k + \bm k'})\mu^4}{5m_{\rm P}^4k^4}(4\mu\mp5k)\,.
\end{eqnarray}
Note that the primary contributions for both integrals come from $p=\mu$. The results above are accordingly insensitive to the infrared cutoff of $p$ and consistent with the assumption $p \gg k$ even though we take $p_{\text{min}}=0$. 
The above result furthermore has an IR singularity, behaving as $\sim 1/k^4$ from small $k$, which can be traced back to the 
infinite ranges of the $t'$ and $t''$ integrals in eq.~(\ref{h_correlation_vac}). It can be explicitly verified that this IR singularity vanishes 
after introducing a finite cutoff to the before-mentioned time integrals. As will be seen in the following section, such a singularity 
is indeed absent in the expanding universe case, where all time integral ranges are finite.

In the general case with finite $T$ and $\mu$, we find
\begin{eqnarray}\nonumber
	\Pi_{+}(\bm p, \theta;\bm k,\lambda)f_{|{\bm k - \bm p}|-\mu}&\approx& \frac{4p^2}{k^4}f_{p-\mu}-\frac{4p}{k^3}\left(\lambda +\cos\theta\left[1-(1-f_{p-\mu})\frac{p}{T}\right]\right) f_{p-\mu}\,,
	\\
	\Pi_{-}(\bm p, \theta;\bm k,\lambda)f_{p+\mu}&\approx&\frac{4p^2}{k^4}f_{p+\mu}+\frac{4p}{k^3}(\lambda-\cos\theta)f_{p+\mu}\,,
\end{eqnarray}
up to $\mathcal{O}(k^{-3})$ when $k \ll T, \mu$. 
From the definition of eq.~(\ref{eq:Stokes_circular}), we can obtain the Stokes parameter $V(\bm{k})$ as
\begin{eqnarray}\nonumber
\label{V_flat}
V(\bm{k}) 
&\approx&\frac{\pi}{2m_{\rm P}^4}\int^{\infty}_{0} {\rm d}p\int^{\pi}_0 {\rm d}\theta p^2\sin\theta
\frac{8p}{k^3}(f_{p+\mu}-f_{p-\mu})	
\\
&=&-\frac{48\pi T^4}{m_{\rm P}^4k^3}\Big[\rm{Li}_{4}\bigl(-{\rm e}^{-\mu/T} \bigr)-\rm{Li}_{4}\bigl(-{\rm e}^{\mu/T}\bigr)\Big],
\end{eqnarray}
where we used
\begin{eqnarray}
	\int^{\infty}_{0} {\rm d}pp^3f_{p-\mu}=-6T^4\rm{Li}_{4}\bigl(-{\rm e}^{\mu/T}\bigr)
\end{eqnarray}
with $\mathrm{Li}_{s}(z) = \sum_{k=1}^{\infty} z^k/k^s$ being the polylogarithm function.
When $\mu\ll T$, eq.~(\ref{V_flat}) reduces to 
\begin{eqnarray}
V(\bm k) \approx -\frac{72\pi T^4}{m_{\rm P}^4k^3}\zeta(3) \frac{\mu}{T} +\mathcal{O}\left(\left(\frac{\mu}{T}\right)^2\right),
\end{eqnarray}
where $\zeta(x)$ is the Riemann zeta function with $\zeta(3) \simeq 1.202$. 
Accordingly, one finds that $V(\bm k)$ vanishes for the $\Pi_{\pm}(\bm p, \theta;\bm k,\lambda)$ terms when $\mu=0$. Even when including the higher-order corrections up to $\mathcal{O}(k^{0})$, this conclusion remains unchanged. 

We, however, find a non-vanishing $V(\bm k)$ for $\mu=0$ by including terms up to $\mathcal{O}(k)$. The nonzero contribution here comes only from the $\Pi_{0}(\bm p, \theta;\bm k,\lambda)$ term, as
\begin{eqnarray}
	V(\bm k) \approx-\frac{\pi k}{15m_{\rm P}^4}\, ,
\end{eqnarray}
which shows that $V(\bm k)$ can be non-vanishing even for $\mu = 0$. 
A detailed analysis going beyond the long wavelength limit is, however, out of the scope of the present study.

\subsection{Expanding universe}
\label{sec:expanding}
In this section, we consider the expanding universe in the radiation dominated period. The scale factor $a(\tau)$ in this period behaves as 
\begin{eqnarray}
a(\tau) = a_{\rm i} \frac{\tau}{\tau_{\rm i}}\,, 
\label{eq:scale_factor}
\end{eqnarray}
where $\tau_{\rm i}$ stands for the initial time at which thermalized left-handed neutrinos are realized and $a_{\rm i}$ is the corresponding scale factor. Furthermore, the solution of eq.\,(\ref{eq:GF}) is known to be%
\footnote{Note that the solution is distinct from the one in ref.~\cite{Anber:2016yqr} due to the different periods considered in the early Universe, which give rise to the different forms of $\mathcal{H}$ in eq.\,(\ref{eq:GF}). Here we have $\mathcal{H}=1/\tau$ for the radiation dominated period as opposed to $\mathcal{H}=-1/\tau$ for the end of inflation considered in ref.~\cite{Anber:2016yqr}. }
\begin{eqnarray}
G(\tau,\tau')=\Theta(\tau-\tau')\tau'\frac{\sin k(\tau-\tau')}{k\tau}\,.
\label{eq:G_solution}
\end{eqnarray}
With these ingredients, we next need to evaluate the $\tau'$, $\tau''$ and $\bm{p}$ integrals in eq.\,(\ref{eq:hcor_th}). Unlike the case in the Minkowski spacetime, it is not possible to compute them analytically for arbitrary $\tau$, $\bm{k}$, $\bar T = a T $ and $\bar \mu = a \mu$ (where these dimensionful quantities are rescaled by the scale factor $a$ as we mention in subsection~\ref{sec:pol_thermal}), while analytical expressions can be obtained for certain physically relevant limits. Specifically, we will consider the regime $\bar T \gg k, \bar \mu$. It will, however, become clear below that both $k$ and $\bar \mu$ need to be non-zero to obtain a non-zero $V$ parameter. Furthermore, $\bar T$ is considered large enough such that $\tau$, $\tau_{\rm i} \gg 1/\bar T$ (see appendix~\ref{sec:estimate_tau}). As will be demonstrated later, these are reasonable assumptions for the temperature ranges of the radiation dominated period considered in this work.

Let us start with the $\tau'$ and $\tau''$ integrals. We define
\begin{eqnarray}
\bar{\mathcal{I}}(\tau,\tau_{\rm i},b) = \int_{\tau_{\rm i}}^{\infty} {\rm d} \tau' \frac{G(\tau,\tau')}{a^2(\tau')} {\rm e}^{{\rm i}b\tau'}, 
\label{eq:eq1}
\end{eqnarray}
where
\begin{eqnarray}
b = |\bm{k} - \bm{p}| - p \equiv b_-
\end{eqnarray}
for the $\Gamma_{\pm}$ terms and 
\begin{eqnarray}
b = |\bm{k} - \bm{p}| + p \equiv b_+
\end{eqnarray}
for the $\Gamma_{0}$ term in eq.\,(\ref{eq:hcor_th}). 
We will treat each case separately in the small $k$ limit and 
keep terms linear in $k$. 
Note that the $\tau''$ integral in eq.\,(\ref{eq:hcor_th}) corresponds to $\bar{\mathcal{I}}(\tau,\tau_{\rm i},-b) = \bar{\mathcal{I}}^{\ast}(\tau,\tau_{\rm i},b)$. 
Hence, we have 
\begin{eqnarray}
\int_{\tau_{\rm i}}^{\infty} {\rm d} \tau' {\rm d} \tau''  \frac{G(\tau,\tau')G(\tau,\tau'')}{a^2(\tau') a^2(\tau'')} {\rm e}^{{\rm i}b\tau'} {\rm e}^{-{\rm i}b\tau''} =  |\bar{\mathcal{I}}(\tau,\tau_{\rm i},b)|^2 = |\bar{\mathcal{I}}(\tau,\tau_{\rm i},-b)|^2, 
\label{eq:eq6}
\end{eqnarray}
and it is therefore sufficient to consider either the $\tau'$ or $\tau''$ integral. The correlation function is furthermore guaranteed to be real. 
Leaving the details of the calculation to appendix \ref{sec:tau_integrals}, we here directly show the final results for the $\Gamma_{\pm}$ and $\Gamma_{0}$ terms.
For $b = b_-$ ($\Gamma_{\pm}$ terms) we have
\begin{eqnarray}
|\bar{\mathcal{I}}(\tau,\tau_{\rm i},b_-)|^2 \simeq 
\frac{\tau_{\rm i}^4}{a_{\rm i}^4} \left[
\log \Bigl( \frac{\tau}{\tau_{\rm i}} \Bigr) - \frac{\tau - \tau_{\rm i}}{\tau} 
\right]^2 
+ \mathcal{O}(k^2)\,. 
\label{eq:eq12}
\end{eqnarray}
For $b = b_+$ ($\Gamma_{0}$ term), the result reads
\begin{eqnarray}
|\bar{\mathcal{I}}(\tau,\tau_{\rm i},b_+)|^2\simeq &&  
\frac{\tau_{\rm i}^4}{a_{\rm i}^4} \Biggl \{ 
\Bigl( \mathrm{ci}(2\tau p) - \mathrm{ci}(2\tau_{\rm i} p) - \frac{1}{\tau p} \sin [(\tau - \tau_{\rm i}) p]  \cos [(\tau + \tau_{\rm i}) p] \Bigr)^2 \nonumber \\
&&+\Bigl( \mathrm{Si}(2\tau p) - \mathrm{Si}(2\tau_{\rm i} p) - \frac{1}{\tau p} \sin [(\tau - \tau_{\rm i}) p] \sin [(\tau + \tau_{\rm i}) p] \Bigr)^2  \nonumber \\
&& + 2\frac{k}{ p} \cos \theta  \sin [(\tau - \tau_{\rm i}) p] 
\Bigr(
\sin [(\tau + \tau_{\rm i}) p] 
\bigl[
\mathrm{ci}(2\tau p) - \mathrm{ci}(2\tau_{\rm i} p) \bigr]  \nonumber \\
&&- \cos [(\tau + \tau_{\rm i}) p] 
\bigl[
\mathrm{Si}(2\tau p) - \mathrm{Si}(2\tau_{\rm i} p) \bigr] \Bigr)
\Biggr \} + \mathcal{O}(k^2)\,,  
\label{eq:eq21}
\end{eqnarray}
where $\mathrm{ci}(x)$ and $\mathrm{Si}(x)$ are the cosine and sine integral functions defined as 
\begin{align}
\mathrm{ci}(x) &\equiv -  \int_x^{\infty} {\rm d}t \frac{\cos t}{t} = \gamma_{\,\mathrm{E}} + \log x + \int_0^x {\rm d}t \frac{\cos t - 1}{t}\,,  
\label{eq:ci}
\\
\mathrm{Si}(x) & \equiv  \int_0^x {\rm d} t \frac{\sin t}{t}\,,
\label{eq:Si}
\end{align}
respectively, with $\gamma_{\,\mathrm{E}}$ standing for the Euler-Mascheroni constant.

Next, we consider the angular integrals of $\bm{p}$, $\phi$ and $\theta$. While the $\phi$ integral is trivial, the $\theta$ integral requires some work, 
but can nevertheless be done analytically in the small $k$ limit. Relegating the details of this calculation to appendix \ref{sec:angular_integrals}, 
we here give the final results individually for the contributions of both $\Gamma_{\pm}$ and $\Gamma_{0}$ terms to eq.\,(\ref{eq:hcor_th}). 
For the $\Gamma_{\pm}$ terms, we have 
\begin{eqnarray}
&&\langle h_{\lambda}(\bm{k}, \tau) h_{\lambda'}(\bm{k}', \tau) \rangle_{\mathrm{th},\,\Gamma_{\pm}} \nonumber \\
&&\simeq  
\frac{32 \pi}{15} \delta_{\lambda,\lambda'}\frac{\delta(\bm{k} + \bm{k}')}{m_{\rm P}^4} \frac{\tau_{\rm i}^4}{a_{\rm i}^4} 
\left[
\log \Bigl( \frac{\tau}{\tau_{\rm i}} \Bigr) - \frac{\tau - \tau_{\rm i}}{\tau} 
\right]^2 \int_0^{\infty} {\rm d}p \ p^4 
\left(1 \mp \lambda \frac{k}{ p} \right)
f_{p \mp \bar \mu} 
 + \mathcal{O}(k^2).   
\label{eq:eq26}
\end{eqnarray}
The expression of the $\Gamma_{0}$ term turns out to be much lengthier. We therefore give the leading-order (LO) and next-to-leading-order (NLO) terms in terms of $k$ 
separately. They read 
\begin{eqnarray}
&& \langle h_{\lambda}(\bm{k}, \tau) h_{\lambda'}(\bm{k}', \tau) \rangle_{\mathrm{th},\,\Gamma_{0},\,\mathrm{LO}} \nonumber \\
&& \simeq
\frac{16 \pi}{5} \delta_{\lambda,\lambda'}\frac{\delta(\bm{k} + \bm{k}')}{m_{\rm P}^4} \frac{\tau_{\rm i}^4}{a_{\rm i}^4} 
\int_0^{\infty} {\rm d}p \ p^4 (f_{p - \bar \mu} + f_{p + \bar \mu}) \nonumber \\
&& \quad\times 
\Biggl \{ 
\Bigl( \mathrm{ci}(2\tau p) - \mathrm{ci}(2\tau_{\rm i} p) - \frac{1}{\tau p} \sin [(\tau - \tau_{\rm i}) p]  \cos [(\tau + \tau_{\rm i}) p] \Bigr)^2 \nonumber \\
&&\quad +
\Bigl( \mathrm{Si}(2\tau p) - \mathrm{Si}(2\tau_{\rm i} p) - \frac{1}{\tau p} \sin [(\tau - \tau_{\rm i}) p] \sin [(\tau + \tau_{\rm i}) p] \Bigr)^2 \Biggr \}\,, 
\label{eq:eq32}
\end{eqnarray}
and 
\begin{eqnarray}
&& \langle h_{\lambda}(\bm{k}, \tau) h_{\lambda'}(\bm{k}', \tau) \rangle_{\mathrm{th},\,\Gamma_{0},\,\mathrm{NLO}} \nonumber \\
&&\simeq  
-\frac{16 \pi}{15} \delta_{\lambda,\lambda'}\frac{\delta(\bm{k} + \bm{k}')}{m_{\rm P}^4} \frac{\tau_{\rm i}^4}{a_{\rm i}^4} \lambda k 
\int_0^{\infty} {\rm d}p \ p^4  \nonumber \\
&&\quad\times \Biggl \{ \frac{2}{p} (f_{p - \bar \mu} + f_{p + \bar \mu}) \sin [(\tau - \tau_{\rm i}) p] 
\Bigr(
\sin [(\tau + \tau_{\rm i}) p] 
\bigl[
\mathrm{ci}(2\tau p) - \mathrm{ci}(2\tau_{\rm i} p) \bigr]  \nonumber \\
&&\quad- \cos [(\tau + \tau_{\rm i}) p] 
\bigl[
\mathrm{Si}(2\tau p) - \mathrm{Si}(2\tau_{\rm i} p) \bigr] \Bigr) \nonumber \\
&&\quad 
+ \left[- \frac{2}{p} (f_{p - \bar \mu} + f_{p + \bar \mu}) + \frac{1}{T} f_{p + \bar \mu}  (1 - f_{p + \bar \mu}) \right] \nonumber \\ 
&&\quad \times \biggl[ \Bigl(
\mathrm{ci}(2\tau p) - \mathrm{ci}(2\tau_{\rm i} p) - \frac{1}{\tau p} \sin [(\tau - \tau_{\rm i}) p]  \cos [(\tau + \tau_{\rm i}) p] \Bigr)^2 \nonumber \\
&&\quad+\Bigl(
\mathrm{Si}(2\tau p) - \mathrm{Si}(2\tau_{\rm i} p) - \frac{1}{\tau p} \sin [(\tau - \tau_{\rm i}) p] \sin [(\tau + \tau_{\rm i}) p] \Bigr)^2 \biggr] \Biggr \}\,.
\label{eq:eq34.6}
\end{eqnarray}

As the last step, we compute the remaining integral of $p$. 
Let us first consider the $\Gamma_{\pm}$ terms which are simpler than $\Gamma_{0}$.
We leave the calculational details to appendix \ref{sec:p_integrals} and give here the result for the sum of the $\Gamma_{+}$ and $\Gamma_{-}$ terms, which reads
\begin{eqnarray}
&& \langle h_{\lambda}(\bm{k}, \tau) h_{\lambda'}(\bm{k}', \tau) \rangle_{\mathrm{th},\,\Gamma_{+} + \Gamma_{-}}  \nonumber \\
&&\simeq -\frac{256 \pi}{5} \delta_{\lambda,\lambda'}\frac{\delta(\bm{k} + \bm{k}')}{m_{\rm P}^4} \frac{\tau_{\rm i}^4}{a_{\rm i}^4} 
\left[ \log \Bigl( \frac{\tau}{\tau_{\rm i}} \Bigr) - \frac{\tau - \tau_{\rm i}}{\tau} \right]^2 \bar T^5  \nonumber \\
&& \quad\times
\Biggl\{  
\mathrm{Li}_{5} \bigl(-{\rm e}^{\bar \mu/\bar T}\bigr) +  \mathrm{Li}_{5} \bigl(-{\rm e}^{-\bar \mu/\bar T}\bigr)  
- \frac{\lambda}{4} \frac{k}{\bar T} \Bigl[ \mathrm{Li}_{4} \bigl(-{\rm e}^{\bar \mu/\bar T}\bigr) - \mathrm{Li}_{4} \bigl(-{\rm e}^{-\bar \mu/\bar T}\bigr) \Bigr] 
\Biggr\} \nonumber \\
&& \quad+ \mathcal{O}\Bigl(\Bigl(\frac{k}{\bar T}\Bigr)^2 \Bigr).
\label{eq:eq52.1}
\end{eqnarray}
Expanding the above expression in $\bar \mu/\bar T$, we have
\begin{eqnarray}
&&\langle h_{\lambda}(\bm{k}, \tau) h_{\lambda'}(\bm{k}', \tau) \rangle_{\mathrm{th},\,\Gamma_{+} + \Gamma_{-}} \nonumber \\
&&\simeq
\frac{96 \pi}{5} \delta_{\lambda,\lambda'}\frac{\delta(\bm{k} + \bm{k}')}{m_{\rm P}^4} \frac{\tau_{\rm i}^4}{a_{\rm i}^4} 
\left[
\log \Bigl( \frac{\tau}{\tau_{\rm i}} \Bigr) - \frac{\tau - \tau_{\rm i}}{\tau} 
\right]^2 \bar T^5 
\left[ 5 \zeta(5) - \zeta(3) \lambda \frac{k}{\bar T} \frac{\bar \mu}{\bar T} \right] \nonumber \\
&& \quad+ \mathcal{O}\Bigl(\Bigl(\frac{k}{\bar T}\Bigr)^2\mathrm{,}\,\Bigl(\frac{\bar \mu}{\bar T}\Bigr)^2 \Bigr),
\label{eq:eq52.2}
\end{eqnarray}
with $\zeta(5) \simeq 1.037$. 

To compute the $\Gamma_{0}$ term, one needs to integrate eqs.\,(\ref{eq:eq32}) and (\ref{eq:eq34.6}), which does not appear to be possible 
analytically in general. However, considering the time and temperature regions that are relevant for the radiation dominant period, 
some of the functions $\mathrm{Si}(x)$ and $\mathrm{ci}(x)$ appearing in these equations 
can be simplified as follows. 
After the change of integration variable from $p$ to $y = p/\bar T$, the arguments of these functions become either 
$x = 2 \tau {\bar T} y$ or $2 \tau_{\rm i} {\bar T} y$, while the integral can be expected to be dominated by contributions from the region $y\sim 1$. 
Noting that $\tau {\bar T} \gg 1$ and $\tau_{\rm i} {\bar T} \gg 1$ in the radiation dominant period (see appendix~\ref{sec:estimate_tau} for details), 
the functions $\mathrm{Si}(2 \tau {\bar T} y)$, $\mathrm{Si}(2 \tau_{\rm i} {\bar T} y)$, $\mathrm{ci}(2 \tau {\bar T} y)$ 
and $\mathrm{ci}(2 \tau_{\rm i} {\bar T} y)$ can be approximated using their asymptotic forms 
\begin{eqnarray}
\mathrm{Si}(x) & = & \frac{\pi}{2} - \frac{\cos x}{x} + \mathcal{O}\left(\frac{1}{x^2}\right), \label{eq:eq63.1} \\
\mathrm{ci}(x) & = & \frac{\sin x}{x} + \mathcal{O}\left(\frac{1}{x^2}\right).
\label{eq:eq63}
\end{eqnarray}

Within these approximations, the $p$ integral in eqs.\,(\ref{eq:eq32}) and (\ref{eq:eq34.6}) can be computed as (see appendix \ref{sec:p_integrals} for details)
\begin{eqnarray}
&& \langle h_{\lambda}(\bm{k}, \tau) h_{\lambda'}(\bm{k}', \tau) \rangle_{\mathrm{th},\,\Gamma_{0}} \nonumber \\
&&\simeq 
-\frac{8 \pi}{15} \delta_{\lambda,\lambda'}\frac{\delta(\bm{k} + \bm{k}')}{m_{\rm P}^4} \frac{\tau_{\rm i}^4}{a_{\rm i}^4} \frac{\tau - \tau_{\rm i}}{\tau \tau_{\rm i}} \bar T^3 
\Biggl \{3 \frac{\tau - \tau_{\rm i}}{\tau \tau_{\rm i}} \Bigl[\mathrm{Li}_{3} \bigl(-{\rm e}^{\bar \mu/\bar T}\bigr) + \mathrm{Li}_{3} \bigl(-{\rm e}^{-\bar \mu/\bar T}\bigr) \Bigr] \nonumber \\
&&\quad + \lambda \frac{k}{\bar T} 
\Bigl[ \frac{\tau - \tau_{\rm i}}{\tau \tau_{\rm i}} \mathrm{Li}_{2} \bigl(-{\rm e}^{\bar \mu/\bar T}\bigr)
- 2\bar T {\rm e}^{\bar \mu/\bar T} \mathrm{Im} \, \Phi^{\ast}\bigl(-{\rm e}^{\bar \mu/\bar T}, 3, 1 - 2{\rm i}(\tau - \tau_{\rm i}) \bar T\bigr)  \nonumber \\
&& \quad
- 2\bar T {\rm e}^{-\bar \mu/\bar T} \mathrm{Im} \, \Phi^{\ast}\bigl(-{\rm e}^{-\bar \mu/\bar T}, 3, 1 - 2{\rm i}(\tau - \tau_{\rm i}) \bar T\bigr) \Bigr] \Biggr \}  + \mathcal{O}\Bigl(\Bigl(\frac{k}{\bar T}\Bigr)^2 \Bigr)\,,  
\label{eq:eq67}
\end{eqnarray}
where $\Phi^{\ast}(z, s, c)$ is the Lerch transcendent function, defined as 
\begin{eqnarray}
\Phi^{\ast}(z, s, c) \equiv \sum_{k=0}^{\infty} \frac{z^k}{\bigl[(c + k)^2 \bigr]^{s/2}}\,.
\label{eq:eq71}
\end{eqnarray}
Expanding up to first order in $\bar \mu/\bar T$, we have 
\begin{eqnarray}
&& \langle h_{\lambda}(\bm{k}, \tau) h_{\lambda'}(\bm{k}', \tau) \rangle_{\mathrm{th},\,\Gamma_{0}} \nonumber \\
&&\simeq
\frac{4\pi}{15} \delta_{\lambda,\lambda'}\frac{\delta(\bm{k} + \bm{k}')}{m_{\rm P}^4} \frac{\tau_{\rm i}^4}{a_{\rm i}^4} \frac{\tau - \tau_{\rm i}}{\tau \tau_{\rm i}} \bar T^3 
\Biggl \{9 \zeta(3) \frac{\tau - \tau_{\rm i}}{\tau \tau_{\rm i}} \nonumber \\
&&\quad + \lambda \frac{k}{\bar T} 
\left(
\frac{\tau - \tau_{\rm i}}{\tau \tau_{\rm i}} \left[ \frac{\pi^2}{6} + 2 \log(2) \frac{\bar \mu}{\bar T} \right] 
- \frac{1}{2(\tau - \tau_{\rm i})^3 \bar T^2} + \pi^3 \bar T \frac{3 + \cosh \bigl[4\pi(\tau - \tau_{\rm i})\bar T\bigr]}{\sinh^3\bigl[2\pi(\tau - \tau_{\rm i})\bar T\bigr]}
\right) 
\Biggr \} \nonumber \\
&& \quad+ \mathcal{O}\Bigl(\Bigl(\frac{k}{\bar T}\Bigr)^2\mathrm{,}\,\Bigl(\frac{\bar \mu}{\bar T}\Bigr)^2 \Bigr)\,. 
\label{eq:eq67.2}
\end{eqnarray}
It is interesting to observe that the difference between the two helicities here does not vanish even for $\mu=0$, which is different from the situation in the $\Gamma_{\pm}$ terms [see eq.\,(\ref{eq:eq52.2})]. 

Let us now add all terms to get the full correlator, which in the small $k/\bar T$ and small $\bar \mu/\bar T$ expansion 
reads 
\begin{eqnarray}
&& \langle h_{\lambda}(\bm{k}, \tau) h_{\lambda'}(\bm{k}', \tau) \rangle_{\mathrm{th},\,\tau_{\rm i} {\bar T} \gg 1} \nonumber \\
&&\simeq
\frac{8 \pi}{15} \delta_{\lambda,\lambda'}\frac{\delta(\bm{k} + \bm{k}')}{m_{\rm P}^4} 
\frac{\tau_{\rm i}^4}{a_{\rm i}^4} \bar T^5 \Biggl\{
180 \zeta(5) \left[
\log \Bigl( \frac{\tau}{\tau_{\rm i}} \Bigr) - \frac{\tau - \tau_{\rm i}}{\tau} 
\right]^2 + \frac{9}{2} \zeta(3) \frac{(\tau - \tau_{\rm i})^2}{\tau^2 \tau_{\rm i}^2 \bar T^2} \nonumber \\
&&\quad + \lambda \frac{k}{\bar T} \Biggl[ \frac{\pi^2}{12} \frac{(\tau - \tau_{\rm i})^2}{\tau^2 \tau_{\rm i}^2 \bar T^2} 
- \frac{1}{4(\tau - \tau_{\rm i})^2 \tau \tau_{\rm i} \bar T^4} + \frac{\pi^3}{2} \frac{\tau - \tau_{\rm i}}{\tau \tau_{\rm i} \bar T}  \frac{3 + \cosh \bigl[4\pi(\tau - \tau_{\rm i})\bar T\bigr]}{\sinh^3\bigl[2\pi(\tau - \tau_{\rm i})\bar T\bigr]} \nonumber \\
&&\quad+ \frac{\bar \mu}{\bar T} 
\biggl(- 36 \zeta(3) 
\left[ \log \Bigl( \frac{\tau}{\tau_{\rm i}} \Bigr) - \frac{\tau - \tau_{\rm i}}{\tau} \right]^2 
+ \log(2) \frac{(\tau - \tau_{\rm i})^2}{\tau^2 \tau_{\rm i}^2 \bar T^2} \biggr) 
\Biggr] \Biggr \} \nonumber \\
&&\quad + \mathcal{O}\Bigl(\Bigl(\frac{k}{\bar T}\Bigr)^2\mathrm{,}\,\Bigl(\frac{\bar \mu}{\bar T}\Bigr)^2 \Bigr)\,. 
\label{eq:eq68}
\end{eqnarray}
One can further clarify the importance of the different terms by considering the $\tau \gg \tau_{\rm i}$ limit. Furthermore, 
as we concluded in eq.\,(\ref{eq:eq61}), factors such as $1/(\tau \bar T)$ or $1/\sinh^3\bigl[2\pi(\tau - \tau_{\rm i})\bar T\bigr]$ 
are strongly suppressed and $1/(\tau_{\rm i} \bar T)$ can in the limit considered here also be regarded as small.  
The correlator thus becomes 
\begin{eqnarray}\label{hRL_correlation_th}
&& \langle h_{\lambda}(\bm{k}, \tau) h_{\lambda'}(\bm{k}', \tau) \rangle_{\mathrm{th},\,\tau_{\rm i} {\bar T} \gg 1} \nonumber \\
&&\simeq
\frac{8 \pi}{15} \delta_{\lambda,\lambda'}\frac{\delta(\bm{k} + \bm{k}')}{m_{\rm P}^4} 
\frac{\tau_{\rm i}^4}{a_{\rm i}^4} \bar T^5 \Biggl\{180 \zeta(5) \log^2 \Bigl( \frac{\tau}{\tau_{\rm i}} \Bigr) 
+ \lambda \frac{k}{\bar T} \biggl[ \frac{\pi^2}{12} \frac{1}{\tau_{\rm i}^2 \bar T^2} 
- 36 \zeta(3) \frac{\bar \mu}{\bar T} \log^2 \Bigl( \frac{\tau}{\tau_{\rm i}} \Bigr)  \biggr] \Biggr \} \nonumber \\
&&\quad + \mathcal{O}\Bigl(\Bigl(\frac{k}{\bar T}\Bigr)^2\mathrm{,}\,\Bigl(\frac{\bar \mu}{\bar T}\Bigr)^2\Bigr)\,.
\label{eq:eq68.2}
\end{eqnarray}

In terms of the Stokes parameters, the above results can be summarized as
\begin{align}
\label{I_expanding}
I(\bm{k})
&\simeq 
	192 \pi \zeta(5) \frac{1}{m_{\rm P}^4} 
	\frac{\tau_{\rm i}^4}{a_{\rm i}^4} \bar T^5 \log^2 \Bigl( \frac{\tau}{\tau_{\rm i}} \Bigr)\,, 
	\\
\label{V_expanding}
V(\bm{k})
&\simeq 
	\frac{192 \pi}{5} \frac{1}{m_{\rm P}^4} 
	\frac{\tau_{\rm i}^4}{a_{\rm i}^4} \bar T^4k  \bigg[ \frac{\pi^2}{432} \frac{1}{\tau_{\rm i}^2 \bar T^2} 
	- \zeta(3)  \frac{\bar \mu}{\bar T} \log^2 \Bigl( \frac{\tau}{\tau_{\rm i}} \Bigr)  \bigg]\,.
\end{align}
For $V(\bm{k})$ we have kept the leading $1/(\tau_{\rm i} \bar T)^2$ term, 
having the situation of vanishing chemical potential in 
mind, in which case it becomes the leading contribution to the difference between the two helicities. 
We will, however, for simplicity ignore this term in the following discussions, as for currently 
available estimates of $\bar \mu/\bar T$ the second term will be dominant. 

\subsection{Comparisons and discussions}
In general, the leading-order correlation functions of gravitational fields in the Minkowski and expanding universes in the long wavelength limit are different. In the Minkowski spacetime, the correlation functions diverge at $k\rightarrow 0$. From the contributions of neutrinos both in vacuum and in thermal equilibrium, 
the parity-even and parity-odd components of correlation functions become $\mathcal{O}(k^{-4})$ and $\mathcal{O}(k^{-3})$, respectively. On the contrary, the correlation functions in the FLRW background are finite at $k\rightarrow 0$. From neutrinos in thermal equilibrium, the parity-even and parity-odd components of correlation functions emerge at $\mathcal{O}(k^{0})$ and $\mathcal{O}(k)$, respectively. 
Compared with eq.~(\ref{V_flat}) in the Minkowski spacetime, the $V$ parameter in eq.~(\ref{V_expanding}) in the expanding universe is thus much suppressed in the long wavelength regime. Note that both the parity-even and parity-odd components are at $\mathcal{O}(k^0)$ in ref.~\cite{Anber:2016yqr} in the long wavelength limit, while a different mechanism for the helicity imbalance generated in the inflation period is considered therein.

When focusing on the primary contribution from a non-zero chemical potential with $\tau_{\rm i} \bar T\gg 1$ and recast the temperature into physical one at $\tau_{\rm i}$, $T_{\rm i} \equiv T(\tau_{\rm i}) = \bar T/a_{\rm i}$, eq.~(\ref{V_expanding}) becomes
\begin{eqnarray}
    \label{V_expanding_2}
	V(\bm{k})
	\simeq 
	-\frac{192\pi\zeta(3)}{5} \frac{\tau_{\rm i}^4}{m_{\rm P}^4} T_{\rm i}^4 k \mathcal{N}_{\nu} \xi \log^2 \Bigl( \frac{\tau}{\tau_{\rm i}} \Bigr)\,,
\end{eqnarray}  
where $\mathcal{N}_{\nu} = 3$ is the number of neutrinos and we 
define the flavor averaged neutrino degeneracy parameter $\xi$ as 
\begin{eqnarray}
\xi \equiv \frac{1}{\mathcal{N}_{\nu}} \sum_{j} \xi_{j},  
\end{eqnarray} 
where $\xi_{j} \equiv \mu_{j}/T = \bar \mu_{j}/\bar T$, in which the subscript ``$j$'' represents the neutrino flavor. 
Note that $V(\bm{k})$ itself depends on the scale factor through $k = a(\tau) k_{\rm phys}(\tau)$. 
In order to make a direct comparison with observations, we may introduce an observable independent of the scale factor,
\begin{eqnarray}
\Delta\chi ({\bm k})\equiv \frac{{V}(\bm{k})}{{I}(\bm{k})}=\frac{\langle h_{+}(\bm k,t)h_{+}(\bm k',t)\rangle_{\text{th}}-\langle h_{-}(\bm k,t)h_{-}(\bm k',t)\rangle_{\text{th}}}{\langle h_{+}(\bm k,t)h_{+}(\bm k',t)\rangle_{\text{th}}+\langle h_{-}(\bm k,t)h_{-}(\bm k',t)\rangle_{\text{th}}}\,.
\end{eqnarray} 
From eq.~(\ref{hRL_correlation_th}), we find
\begin{eqnarray}\label{Delta_chi_final}
\Delta\chi ({\bm k})=-\frac{\zeta(3)}{5\zeta(5)}\xi \frac{k}{\bar T}=-\frac{\zeta(3)}{5\zeta(5)}\xi \frac{k_{\rm g}}{T_{\rm g}}\,,
\end{eqnarray}
where we take $k_{\rm g}=k_{\rm phys}(\tau_{\rm g})$ and $T_{\rm g}=T(\tau_{\rm g})$ with $k_{\rm phys}(\tau)=k/a(\tau)$ and $\tau_{\rm g}$ the conformal time when gravitational waves are generated from the thermal plasma. Note that $\Delta\chi ({\bm k})$ is not suppressed by $m_{\rm P}$ and depends on three parameters, $k_{\rm g}$, $T_{\rm g}$ and $\xi$;
the parameter $k_{\rm g}$ can be related to the momentum (wavenumber) of gravitational waves measured at present time via $k_{\rm g}=a(\tau_{0}) k_{\rm phys}(\tau_{0})/a(\tau_{\rm g})$, where $\tau_{0}$ denotes the present conformal time and $k_{\rm phy}(\tau_{0})$ corresponds to the measured momentum of gravitational waves; $T_{\rm g}$ may be chosen as $T_{\rm g} \gtrsim 1\,{\rm MeV}$ when neutrinos are still in equilibrium in a primordial plasma during the radiation dominant era; $\xi$ is not well constrained, but at least for electron neutrinos known to be small, the recent ref.~\cite{Matsumoto:2022tlr} for instance obtaining a value of $\xi_{\rm e} = 0.05 \pm 0.03$ (see also the recent ref.~\cite{Seto:2021tad}, which reports a similar result and refs.~\cite{Joyce:1997uy,Serpico:2005bc,Pastor:2008ti,Steigman:2012ve} for earlier estimates). In ref.~\cite{Kawasaki:2022hvx}, a possible theoretical model that could generate a sufficiently large lepton asymmetry consistent with the estimate in ref.~\cite{Matsumoto:2022tlr} is also proposed. Yet the exact value of $\xi$ is still not fully determined from present observations for all neutrino types. In this work, we assume that the chemical potentials of all neutrino flavors are equilibrated and hence $\xi_{\rm e} = \xi_{\mu} = \xi_{\tau} = \xi$ \cite{Steigman:2012ve} and employ the numerical 
value of ref.~\cite{Matsumoto:2022tlr} quoted above.

The computation for $\Delta\chi ({\bm k})$ above only incorporates the gravitational waves produced from thermalized neutrinos. In practice, we should further include other sources that contribute to $I(\bm k)$ and $V(\bm k)$. Considering only the gravitational waves produced by matter in thermal equilibrium from a primordial plasma, we may simply multiply $I(\bm k)$ in eq.~(\ref{I_expanding}) by the effective degrees of freedom. Recall that $I(\bm k)$ is dominated by the $\Gamma_{\pm}$ terms associated with the integral $I_1(k,\pm \bar \mu, \bar T)$ in eq.~(\ref{eq:I1}) for $\mu\ll T$. We hence simply replace the fermionic distribution function in $I_1(k,\pm \bar \mu, \bar T)$ by its bosonic counterpart to roughly estimate the similar bosonic contributions. In addition, we neglect particle masses, assuming that they are much smaller than the considered temperature scales. Consequently, the effective degrees of freedom for $I(\bm k)$ here can be characterized by
\begin{eqnarray}
\mathcal{N}_{\rm d.o.f.}=\mathcal{N}_{\rm f}+\frac{16}{15}\mathcal{N}_{\rm b}\,,
\end{eqnarray}     
due to the relation 
\begin{eqnarray}
	\int_0^{\infty} {\rm d}x
	\frac{x^4}{{\rm e}^{x} -1}=\frac{16}{15}\int_0^{\infty} {\rm d}x 
	\frac{x^4}{{\rm e}^{x}+1}\,,
\end{eqnarray}
where $\mathcal{N}_{\rm f}$ and $\mathcal{N}_{\rm b}$ denote the total numbers of fermions and bosons in thermal equilibrium, respectively. For $V(\bm k)$, the LO contribution arises due to the non-vanishing neutrino degeneracy parameter $\xi$. We eventually obtain a generic expression
\begin{eqnarray}
\label{main}
	\Delta\chi ({\bm k})=-\frac{\zeta(3)r(T_{\rm g}) \mathcal{N}_{\nu}}{5\zeta(5)\mathcal{N}_{\rm d.o.f.}}\frac{k_{\rm phys}(\tau_{0})}{T_{\rm g}}
	\xi\,.
\end{eqnarray}
Here, $r(T_{\rm g})\equiv a(\tau_{0})/a(\tau_{\rm g})$, for which we may utilize \cite{Caprini:2018mtu}
\begin{eqnarray}
\label{eq:r}
r(T_{\rm g})
\simeq 1.25\times 10^{13}\left(\frac{g_{\ast s}(T_{\rm g})}{100}\right)^{1/3}\left(\frac{T_{\rm g}}{\rm GeV}\right)\,,
\label{eq:scale_ratio}
\end{eqnarray} 
where $g_{\ast s}(T_{\rm g})$ represents effective entropic degrees of freedom at $T_{\rm g}$. 
As an example, we choose $T_{\rm g}=200\,{\rm MeV}$ just above the QCD phase transition. Accordingly, we take $\mathcal{N}_{\rm f}=50$, $\mathcal{N}_{\rm b}=18$, and $g_{\ast s}(T_{\rm g})\approx 61.75$ \cite{Husdal:2016haj}, which yield $\mathcal{N}_{\rm d.o.f.}\approx 69.2$ and $r(T_{\rm g})\approx 2.13\times 10^{12}$. In such a case, we obtain
\begin{eqnarray}
\Delta\chi ({\bm k})\approx -5.35\times 10^9\left(\frac{k_{\rm phys}(\tau_{\rm 0})}{\rm GeV}\right)
=-2.21\times 10^{-14}\left(\frac{f}{\rm Hz}\right)\, 
\end{eqnarray} 
by taking $\xi=0.05$, 
where we have defined $f \equiv 2\pi k_{\rm phys}(\tau_{\rm 0})$.

Let us next estimate the gravitational wave spectrum emitted from thermalized primordial neutrinos that can be measured at the present time. 
For this purpose (and for making a direct comparison to the existing literature simpler), we define the dimensionless power spectra $h^2_{I}(\bm k, \eta)$ and 
$h^2_{V}(\bm k, \eta)$ as 
\begin{eqnarray}
\label{eq:dimensionless_power_spec}
\langle h_{\lambda}(\bm k, \eta) h_{\lambda'}^{\ast}(\bm k',\eta)\rangle &=& 
\frac{8\pi^5}{k^3} \delta(\bm k - \bm k') \delta_{\lambda \lambda'}[h^2_{I}(\bm k, \eta) + \lambda h^2_{V}(\bm k, \eta)]\,,  
\end{eqnarray}
where we have 
\begin{eqnarray}
\label{eq:dimensionless_power_spec_I}
h^2_{I}(\bm k, \tau) 
& \simeq &
	\frac{12 \zeta(5)}{\pi^4} \frac{1}{m_{\rm P}^4} 
	\frac{\tau_{\rm i}^4}{a_{\rm i}^4} k^3 \bar T^5 \log^2 \left( \frac{\tau}{\tau_{\rm i}} \right)\,, \\
\label{eq:dimensionless_power_spec_V}
h^2_{V}(\bm k, \tau) 
& \simeq & 
    -\frac{12 \zeta(3)}{5 \pi^4}  \frac{1}{m_{\rm P}^4} 
	\frac{\tau_{\rm i}^4}{a_{\rm i}^4}  k^4 \bar T^4  \xi \log^2 \left( \frac{\tau}{\tau_{\rm i}} \right)\,.
\end{eqnarray}

To express the gravitational wave spectrum as a function of frequency $f$ at the present time, we use \cite{Caprini:2018mtu}
\begin{eqnarray}
\label{eq:Omega_GW_def}
h^2 \Omega^{I/V}_{\rm{GW}}(f) &=& \frac{2 \pi^2}{3} \frac{h^2}{H_0^2} f^2 h_{I/V}^2(f) \nonumber \\
&=& \frac{2 \pi^2}{3} 10^{-4} \left(\frac{\rm{Mpc}}{\rm{km}}\right)^2  \left(\frac{f}{\rm{Hz}}\right)^2 h_{I/V}^2(f)\,,
\end{eqnarray}
where $H_0$ is the Hubble constant today, which is related to the parameter $h$ as 
$H_0 = 100\, h\, \rm{km}\, \rm{s}^{-1}\, \rm{Mpc}^{-1}$. Furthermore, $h_{I/V}^2(f)$ and $f$ are obtained from 
$h^2_{I/V}(\bm k, \tau)$ at time $\tau$ and the momentum $\bm k$ 
as \cite{Turner:1993vb,Nakayama:2008wy,Okano:2020uyr}
\begin{eqnarray}
\label{eq:h_f}
h^2_{I/V}(f) &=& \left(\frac{\Omega_m}{\Omega_{\Lambda}} \right)^2 
\left(\frac{g_{\ast}(T)}{g_{\ast}(T_0)}\right) \left(\frac{g_{\ast s}(T_0)}{g_{\ast s}(T)}\right)^{4/3} 
\overline{\left(\frac{3j_1(k\tau_0)}{k \tau_0}\right)^2} T_1^2\left(\frac{k_{\mathrm{phys}}}{k_{\mathrm{eq}}}\right) h^2_{I/V}(\bm k, \tau)\,,
		\\
\label{eq:f_k_relation}		
f &=& \frac{a(\tau)}{a(\tau_0)} \frac{k_{\mathrm{phys}}(\tau)}{2\pi}\,.
\end{eqnarray}
Here, the factor $\Omega_m/\Omega_{\Lambda}$ expresses the ratio of matter and dark energy in the 
current universe ($\Omega_m = 0.3$, $\Omega_{\Lambda}=0.7$). $g_{\ast}(T)$ denotes, similar to the entropic degrees of freedom $g_{\ast s}(T)$, the effective number of relativistic degrees of freedom contributing to the energy density at temperature $T$. $j_1(x) = (\sin x/x - \cos x)/x$ is the spherical Bessel function and the bar above it 
represents the averaging over multiple periods. 
Considering its argument $x = k\tau_0$, we use $\tau_0 = 1.4 \times 10^3\,{\rm Mpc}$ and $k = k_{\mathrm{phys}}(\tau_0)/a(\tau_0)$ with $a(\tau_0) = 1$ to obtain 
\begin{eqnarray}
k\tau_0 
&=& 9.1 \times 10^{17} \left(\frac{f}{\mathrm{Hz}}\right) \gg 1\,,
\label{eq:ktau0}
\end{eqnarray}
for all frequencies $f$ relevant for realistic measurements. 
Therefore, we can approximate $j_1(x) \simeq -\cos x/x$ and the barred factor becomes 
\begin{eqnarray}
\label{eq:barred}
\overline{\left(\frac{3j_1(k\tau_0)}{k \tau_0}\right)^2} 
&\simeq& \frac{9}{2} \frac{1}{(k \tau_0)^4}\,.
\end{eqnarray}
$T_1(x)$ is the so-called transfer function which describes 
the transfer effect of the universe from the radiation dominant 
to matter dominant 
eras and is obtained by 
numerically solving the equation governing $h_{\lambda}(\bm k, \tau)$ in the expanding universe. 
It can be given as
\begin{eqnarray}
\label{eq:transfer_f}
T^2_1(x) = 1 + 1.57x + 3.42x^2.
\end{eqnarray}
The momentum $k_{\mathrm{eq}}$ appearing in eq.~(\ref{eq:h_f}) is determined by the conformal time $\tau_{\mathrm{eq}}$ corresponding 
to the matter-radiation equality, $k_{\mathrm{eq}} = 1/\tau_{\mathrm{eq}} = 7.1 \times 10^{-2} \Omega_{m} h^2\, \mathrm{Mpc}^{-1}$ with $h = 0.7$. 
Moreover, the ratio $a(\tau)/a(\tau_0)$ appearing in eq.~(\ref{eq:f_k_relation}) can, similar to eq.~(\ref{eq:scale_ratio}), be related to the physical temperatures at times $\tau$ and $\tau_0$ 
from the principle of conservation of entropy, giving 
\begin{eqnarray}
\label{eq:a_tau/a_0}
\frac{a(\tau)}{a(\tau_0)} = \frac{T(\tau_0)}{T(\tau)} \left(\frac{g_{\ast s}(T_0)}{g_{\ast s}(T)}\right)^{1/3}\,.
\end{eqnarray}

It should be noted that eq.~(\ref{eq:h_f}) is only applicable for gravitational waves generated during the radiation dominant period that do not cross the horizon. This leads to a constraint of the $\tau_{\rm i}$ values (or their corresponding temperatures) that can be used for a specific frequency $f$. The respective condition reads 
\begin{eqnarray}
k \tau_{\rm i} > 1. 
\label{eq:horizon_cond}
\end{eqnarray}
To make the physical meaning of this condition more transparent, let us rewrite it in terms of the temperature $T_{\rm i}$ and 
the present day frequency $f$. $k \tau_{\rm i}$ can be obtained by multiplying the factor $\tau_{\rm i}/\tau_0$ to eq.~(\ref{eq:ktau0}). 
We have  
\begin{eqnarray}\label{eq:ratio_tautaui}
\frac{\tau_{\rm i}}{\tau_0} &=& \frac{\tau_{\rm i}}{\tau_{\mathrm{eq}}} \frac{\tau_{\mathrm{eq}}}{\tau_0} = \frac{a(\tau_{\rm i})}{a(\tau_{\mathrm{eq}})} \left( \frac{a(\tau_{\mathrm{eq}})}{a(\tau_0)} \right)^{1/2}\,, \nonumber \\
&=& \left( \frac{g_{\ast s}(T_{\mathrm{eq}}) g_{\ast s}(T_0)}{g_{\ast s}^{2}(T_{\rm i})} \right)^{1/6}
\left( \frac{T_{\mathrm{eq}} T_0}{T^2_{\rm i}} \right)^{1/2} \,, \nonumber \\
&=& 4.6 \times 10^{-12} \left( \frac{T_{\rm i}}{\mathrm{GeV}} \right)^{-1}\,, 
\end{eqnarray}
where $T_{\mathrm{eq}} = 0.8\,{\rm eV}$ is the temperature at the matter-radiation equality with $g_{\ast s}(T_{\mathrm{eq}}) = 3.91$. Here we have used eq.~(\ref{eq:scale_factor}) in the radiation dominated period and $a(\tau) \propto \tau^2$ in the matter dominated period to derive the second equality of the first line in eq.~(\ref{eq:ratio_tautaui}). Furthermore, $T_0 = 2.35 \times 10^{-13}\,{\rm GeV}$, $g_{\ast s}(T_0) = 3.91$ and 
$g_{\ast s}(T_{\rm i}) = 106.75$, the latter implying that 
$T_{\rm i}$ is large enough to thermalize all particles of the standard model.
In combination with 
eq.~(\ref{eq:ktau0}), eq.~(\ref{eq:horizon_cond}) can hence be recast as 
\begin{eqnarray}
4.2 \times 10^{6} \left( \frac{T_{\rm i}}{\mathrm{GeV}} \right)^{-1} \left(\frac{f}{\mathrm{Hz}}\right)  > 1,  
\label{eq:horizon_cond_2}
\end{eqnarray}
which shows that for, say, $f \sim 1\,{\rm Hz}$, $T_{\rm i}$ should not be much larger than $10^6\,{\rm GeV}$. 

Rewriting $h^2_{I/V}(\bm k, \tau)$ in terms of physical quantities, combining eqs.~(\ref{eq:Omega_GW_def})--(\ref{eq:a_tau/a_0}), 
we arrive at%
\footnote{Here we again use eq.~(\ref{eq:scale_factor}) in the radiation dominated period and the relation~(\ref{eq:a_tau/a_0}) based on the entropy conservation to evaluate $\ln^2(\tau/\tau_{\rm i})$. Such an estimation could be different from evaluating $\tau/\tau_{\rm i}$ by using the estimation in eqs.~(\ref{eq:eq61}) and (\ref{eq:tau_i_bar_T_relation_2}) in light of the energy conservation, while their difference in the numerical results is negligible.}
\begin{eqnarray}
\label{eq:spectrum_now_I}
h^2 \Omega_{\rm{GW}}^{I}(f) 
&=& 
\frac{23328 \zeta(5)}{\pi^3} \mathcal{N}_{\nu} \left(\frac{\rm{Mpc}}{\rm{km}}\right)^2  \left(\frac{\rm{Hz}}{\rm{GeV}}\right)^3 
\frac{1}{g_{\ast}(T) g_{\ast}(T_0)} \left(\frac{g_{\ast s}(T_0)}{g_{\ast s}(T)}\right)^{1/3} \left(\frac{T_0}{\rm{GeV}}\right)^{-3} \nonumber \\
&&\times
\left(\frac{\Omega_m}{\Omega_{\Lambda}} \right)^2 (k\tau_0)^{-4}\, T_1^2\left(\frac{k_{\mathrm{phys}}}{k_{\mathrm{eq}}}\right) \log^2 \left[ \frac{T_{\rm i}}{T}  \left(\frac{g_{\ast s}(T_{\rm i})}{g_{\ast s}(T)}\right)^{1/3} \right] \left(\frac{f}{\rm{Hz}}\right)^5\,,  \\
\label{eq:spectrum_now_V}
h^2 \Omega_{\rm{GW}}^{V}(f) 
&=& 
\frac{46656 \zeta(3)}{5 \pi^2} \mathcal{N}_{\nu} \left(\frac{\rm{Mpc}}{\rm{km}}\right)^2  \left(\frac{\rm{Hz}}{\rm{GeV}}\right)^4 
\frac{1}{g_{\ast}(T) g_{\ast}(T_0)} \left(\frac{T_0}{\rm{GeV}}\right)^{-4} \nonumber \\
&&\times
\left(\frac{\Omega_m}{\Omega_{\Lambda}} \right)^2 (k\tau_0)^{-4}\, T_1^2\left(\frac{k_{\mathrm{phys}}}{k_{\mathrm{eq}}}\right) \xi 
\log^2 \left[ \frac{T_{\rm i}}{T}  \left(\frac{g_{\ast s}(T_{\rm i})}{g_{\ast s}(T)}\right)^{1/3} \right]
\left(\frac{f}{\rm{Hz}}\right)^6\,. 
\end{eqnarray}

For making a numerical estimate, we need to fix $\tau$, the respective temperature $T$ and the corresponding active degrees of freedom. 
While keeping $\tau_{\rm i}$ and $T_{\rm i}$ as free parameters (with $g_{\ast}(T_{\rm i}) = g_{\ast s}(T_{\rm i})  = 106.75$ as mentioned above), 
$\tau$ in our approach marks the end of the period during which neutrinos emit gravitational waves, for which we take $T(\tau) = 1\,{\rm MeV}$ at  
the neutrino decoupling temperature and $g_{\ast}(T) = g_{\ast s}(T) = 14.25$. 
We note that neutrinos will in principle continue to generate gravitational waves even after $\tau$. For times between the neutrino decoupling and the end of the radiation dominant period, the thermal distribution of neutrinos is frozen, with momenta that are rescaled due to the expanding universe. This rescaling allows one to define a time-dependent temperature analogous to the situation before $\tau$. 
The rescaled momenta $\bm k$ and $\bm p$ in eq.~(\ref{eq:hcor_th}) and $\bar T$, can however not both be considered as constants and our analytical results derived above can therefore no longer be applied. As the universe enters the matter dominant era, $a(\tau)$ can furthermore no longer be described by eq.~(\ref{eq:scale_factor}). 

Substituting the above numerical values, $\xi=0.05$, and  $g_{\ast}(T_0) = 2$, we have
\begin{eqnarray}
\label{eq:spectrum_now_I_tau_1}
h^2 \Omega_{\rm{GW}}^{I}(f) 
&=& 
1.4 \times 10^{-32} \times \left[1 + \frac{1}{49.9} \log \left(\frac{T_{\rm i}}{m_{\mathrm{P}}}\right)\right]^2\left(\frac{f}{\rm{Hz}}\right)^3\,, \\
\label{eq:spectrum_now_V_tau_1}
h^2 \Omega_{\rm{GW}}^{V}(f) 
&=& 
2.9 \times 10^{-44} \times \left[1 + \frac{1}{49.9} \log \left(\frac{T_{\rm i}}{m_{\mathrm{P}}}\right)\right]^2\left(\frac{f}{\rm{Hz}}\right)^4\,,
\end{eqnarray}
which shows that 
the parity-odd component is (at $f = 1\,\mathrm{Hz}$) suppressed by more than 11 orders of magnitude compared to its 
parity-even counterpart. This is caused primarily by one additional factor of $k/\bar{T}$ in its power spectrum 
and secondarily by the further suppression due to the smallness of $\xi$. The parity-even gravitational spectrum due to hydrodynamic fluctuations of the primordial plasma has been previously obtained from the emission rate of gravitons based on a perturbative Standard Model calculation \cite{Ghiglieri:2015nfa,Ghiglieri:2020mhm,Ringwald:2020ist}. 

\begin{figure}
\centering
\includegraphics[width=0.75\linewidth]{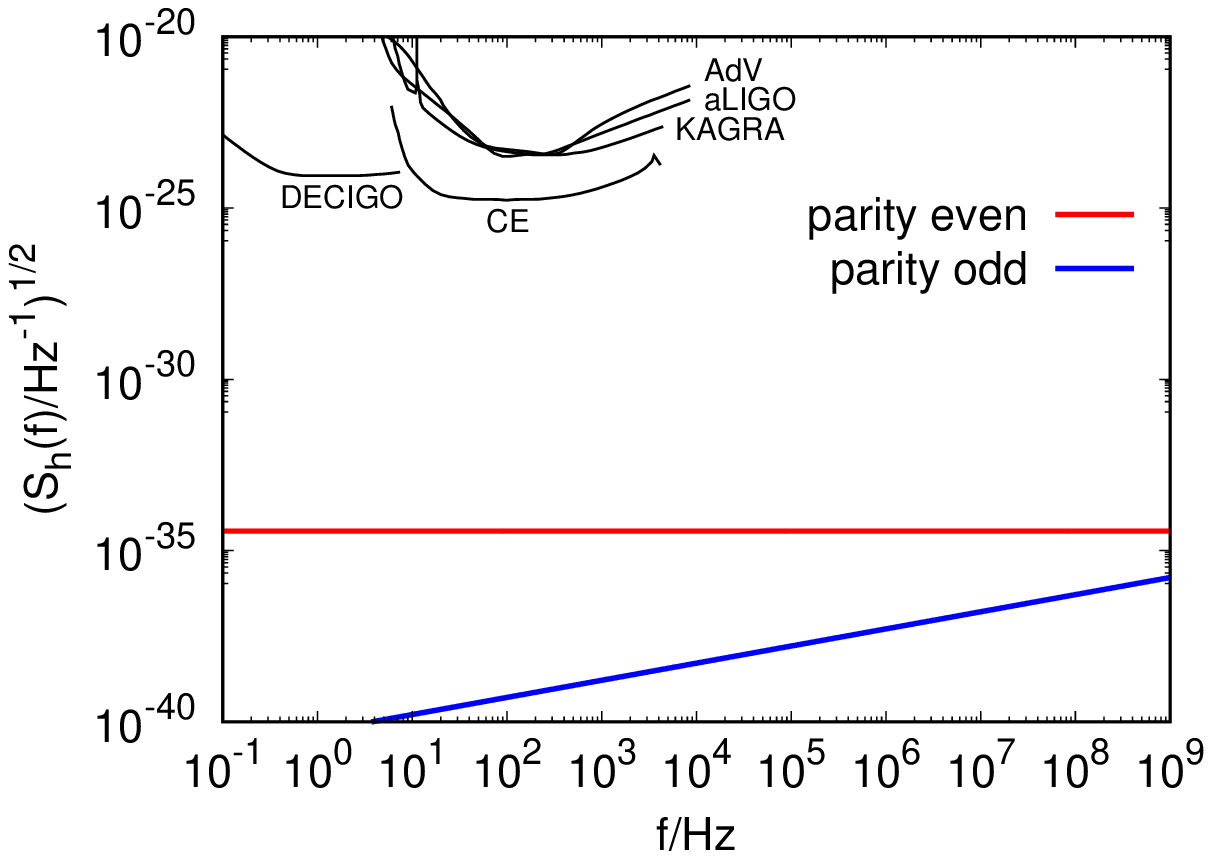}
\includegraphics[width=0.75\linewidth]{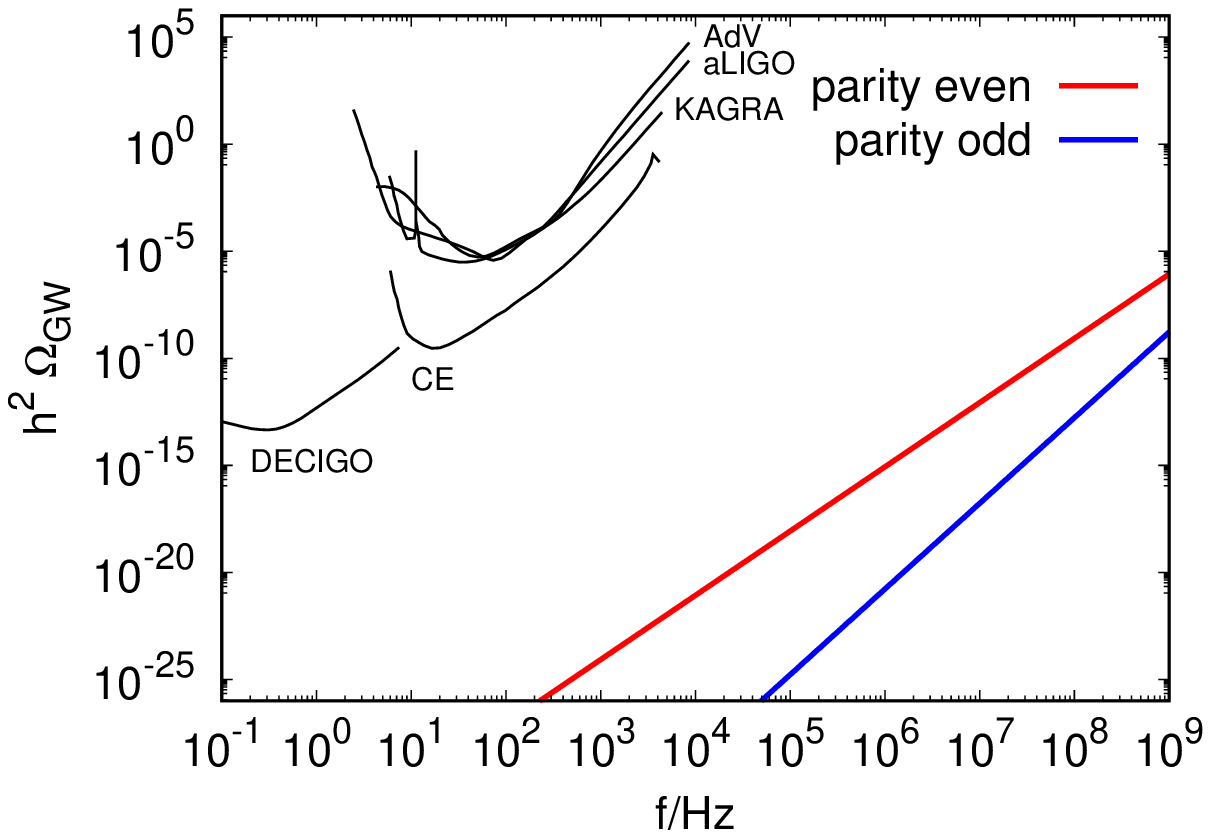}
\caption{The power spectral densities of the gravitational waves generated from primordial thermal neutrinos. The even (odd) parity 
component is shown by a thick red (blue) line. For comparison, the sensitivity curves of various GW detectors, extracted from ref.~\cite{GWplotter}, are shown as thin black lines.}
\label{fig:sensitivity_comparison}
\end{figure}

Another frequently used measure to quantify the strength of a gravitational wave signal is the 
power spectral density $S_h(f)$ \cite{Moore:2014lga}, which has the dimension of $\rm{Hz}^{-1}$ and is related to 
$h^2 \Omega_{\rm{GW}}(f)$ as 
\begin{eqnarray}
\label{eq:power_spectral_density}
\frac{S_h(f)}{\rm{Hz}^{-1}} = \frac{3}{2\pi^2} 10^4 \left(\frac{\rm{Mpc}}{\rm{km}}\right)^{-2} \left(\frac{f}{\rm{Hz}}\right)^{-3} h^2 \Omega_{\rm{GW}}(f).
\end{eqnarray}
In figure~\ref{fig:sensitivity_comparison}, we show $\sqrt{S_h(f)/\rm{Hz}^{-1}}$ and $h^2 \Omega_{\rm{GW}}(f)$ as functions of the frequency $f/\rm{Hz}$ together with the sensitivity curves of a few GW detectors \cite{Moore:2014lga,GWplotter}. 
$T_{\rm i} = 10^2\,{\rm GeV}$ was used in this figure, corresponding to the temperature just below the electroweak (EW) phase transition \cite{DOnofrio:2015gop}. For temperatures above the EW phase transition, there may be further chiral 
asymmetries related to different conserved charges (see, for instance, ref.\,\cite{Domcke:2020quw}), which will 
inevitably be model dependent. We however note that the results of this paper can be easily generalized to any model 
before the EW phase transition by summing up all chiral asymmetries related to its available chemical potentials.
As seen in the plots of figure~\ref{fig:sensitivity_comparison}, 
the GW signals of thermal neutrinos are likely too weak to be measured in the foreseeable future. 
Let us, however, mention here the possibility of measuring high frequency GWs 
(see refs.~\cite{Cruise:2006zt,Akutsu:2008qv,Cruise:2012zz,Ito:2019wcb,Ejlli:2019bqj} for recent developments in this 
direction), which might be sensitive to the signals discussed in this work. To make more accurate statements about this 
possibility, one would however need to go beyond the small $k/\bar{T}$ expansion used here to study the behavior 
of $h^2 \Omega_{\rm{GW}}(f)$ in the high frequency limit.

As a last point, we check the validity of the small $k/\bar{T}$ expansion, that we have used to derive our analytic results 
in view of the experimentally measurable frequency ranges illustrated in figure~\ref{fig:sensitivity_comparison}. 
As shown in this plot, typical operating or planned GW detectors 
can measure frequency ranges up to about $10^4\,{\rm Hz}$. Let us translate this into the corresponding range of $k/\bar{T}$. 
Using eq.~(\ref{eq:f_k_relation}), we have 
\begin{eqnarray}
\label{eq:k_T_ratio}
\frac{k}{\bar{T}} &=& 2\pi \frac{a(\tau_0)}{a(\tau)} \frac{f}{T} \nonumber \\
&=& 2 \pi \left(\frac{\rm{Hz}}{\rm{GeV}}\right) \left(\frac{T(\tau_0)}{\rm{GeV}}\right)^{-1} \left(\frac{g_{\ast s}(T)}{g_{\ast s}(T_0)}\right)^{1/3} 
\left(\frac{f}{\rm{Hz}}\right) \nonumber \\
&=& 1.8 \times 10^{-11} \left(\frac{g_{\ast s}(T)}{g_{\ast s}(T_0)}\right)^{1/3} \left(\frac{f}{\rm{Hz}}\right), 
\end{eqnarray}
where we have in the second line used eq.~(\ref{eq:a_tau/a_0}). Due to the factor involving the degrees of freedom $g_{\ast s}(T)$, 
the coefficient of $f/\rm{Hz}$ weakly depends on $\tau$. This factor takes values ranging from about 3 above $200\,{\rm GeV}$, down to 1 at eV scales. 
For whatever time or temperature is considered, the ratio $k/\bar{T}$ is small for frequencies $f \lesssim 10^{10}\,{\rm Hz}$. 
Hence, for frequencies shown in figure~\ref{fig:sensitivity_comparison} this requirement is satisfied, which shows that the small $k/\bar{T}$ expansion 
is valid. 
The results of this work do not allow us to make predictions about what happens for frequencies of the order of $f \sim 10^{11}\,{\rm Hz}$ or larger due to the adopted approximation. 
It is, however, possible to make the educated guess that because of the thermal distribution factors appearing in 
eq.~(\ref{eq:hcor_th}), the GW spectrum will be exponentially suppressed roughly as ${\rm e}^{-k/\bar{T}}$ in the high frequency limit. This exponential suppression can be expected to become dominant for $f \gtrsim 10^{11}\,{\rm Hz}$. 
We note that a 
similar behavior was obtained in the explicit calculations of refs.~\cite{Ghiglieri:2015nfa,Ringwald:2020ist}.

\section{Summary and outlook}
\label{sec:summary}
We have studied the birefringence of gravitational waves triggered by primordial left-handed neutrinos in thermal equilibrium. A generic expression for correlation functions of gravitational fluctuations with different helicities due to chiral fermions at finite temperature and density is derived and further applied to analyze the Stokes parameters of gravitational waves induced by thermalized neutrinos in the early Universe. We investigated both the cases of Minkowski and expanding spacetime backgrounds and obtained a non-vanishing $V$ parameter characterizing gravitational birefringence for non-zero neutrino chemical potentials. In the expanding universe, the general expression for the parity-violating parameter $\Delta \chi$ is given by eq.~(\ref{main}).

While we have in this work focused on the thermalized neutrinos in the early Universe, our argument and derivation may be extended to the thermalized neutrino as chiral matter in core-collapse supernovae \cite{Yamamoto:2015gzz}, which is expected to produce polarized gravitational waves as well. More generically, chiral gravitational waves from neutrinos or other chiral fermions can also be generated in out of equilibrium situations both in the early Universe and supernovae. To study such situations, one needs to use the radiation transport theory taking into account the effects of chirality of fermions, termed the chiral radiation transport theory \cite{Yamamoto:2020zrs}, in which the energy-momentum tensor with chiral corrections is expressed in terms of the distribution function of chiral fermions. These questions are deferred to future work.

\acknowledgments
We thank Kohei Kamada, Jun'ya Kume, Kin-Wang Ng, and Yusuke Yamada for useful discussions and comments. 
P.G. is supported by the Grant-in Aid for Scientific Research (C) (JSPS KAKENHI Grant Number JP20K03940) and the Leading Initiative for Excellent Young Researchers (LEADER) of the Japan Society for the Promotion of Science (JSPS).
N.Y. is supported by the Keio Institute of Pure and Applied Sciences 
(KiPAS) project at Keio University and JSPS KAKENHI Grant Number JP19K03852.
D.-L.Y. is supported by the National Science and Technology Council (Taiwan) under Grant Number MOST 110-2112-M-001-070-MY3.

\appendix
\section{Derivation of eqs.\,(\ref{eq:hcor_th})--(\ref{eq:Gamma_bar_0})}
\label{sec:app_thermal_part}
In this appendix, we demonstrate all necessary steps needed to obtain eqs.\,(\ref{eq:hcor_th})--(\ref{eq:Gamma_bar_0}). 
For simplicity of notation, we here perform computations without rescaling dimensionful quantities with respect to the scaling factor $a$ (which can be done later). 
We start from eq.\,(\ref{Mlambda}) and replace the vacuum expectation value with its thermal counterpart of eq.\,(\ref{eq:thermal_equib}). 
First, let us define 
\begin{eqnarray}\nonumber
	\hat{M}_{\lambda\lambda'} &\equiv& L^{\dagger}({\bm p - \bm k},\tau')\epsilon^{\mu\nu}_{-\lambda}(\bm k)\big[\bar{\sigma}_{\mu}p_{\nu}+\bar{\sigma}_{\mu}(p-k)_{\nu}\big] 
	L(\bm p, \tau') \\
	&& \times L^{\dagger}({\bm p' - \bm k'},\tau'')\epsilon^{\rho\sigma}_{-\lambda'}(\bm k')
	\big[\bar{\sigma}_{\rho}p'_{\sigma}+\bar{\sigma}_{\rho}(p'-k')_{\sigma}\big]L(\bm p',\tau''), 
\end{eqnarray}
and compute $\Tr(\hat{M}_{\lambda\lambda'} \hat{\rho})$ according to eqs.\,(\ref{eq:thermal_equib}) and (\ref{eq:trace}). 
Throughout the following discussion, trivial terms that lead to 
$\bm k=\bm k'= \bm 0$ will be neglected. 

Considering first the pure vacuum term in the trace of eq.\,(\ref{eq:trace}), $\langle 0|\hat{M}_{\lambda\lambda'} \hat{\rho} |0 \rangle$, the non-vanishing operator combination reads
\begin{eqnarray}\label{gstate_comb}
\Omega_{0}\equiv\langle 0|b_{\bm k - \bm p}a_{\bm p}a^{\dagger}_{\bm p' - \bm k'}b^{\dagger}_{-\bm p'}\hat{\rho} |0 \rangle
=(2\pi)^6\delta^3({\bm k + \bm k'})\delta^3({\bm k - \bm p + \bm p'}) \frac{1}{Z} {\rm e}^{\beta (E_q+\mu)}\,.
\end{eqnarray}
Next, for $\langle 0|a_{\bm q}\hat{M}_{\lambda\lambda'}\hat{\rho} a_{\bm q}^{\dagger}|0 \rangle$, the non-trivial terms are given by
\begin{eqnarray}
\Omega^{\text{th}(1)}_{a} &\equiv& \langle 0|a_{\bm q}a^{\dagger}_{\bm p - \bm k}a_{\bm p}a^{\dagger}_{\bm p' - \bm k'}a_{\bm p'}\hat{\rho} a^{\dagger}_{\bm q}|0 \rangle \nonumber \\
&=&(2\pi)^9\delta^3({\bm k + \bm k'})\delta^3({\bm p'- \bm p + \bm k})\delta^3({\bm q - \bm p'}) \frac{1}{Z} {\rm e}^{\beta\left[(E_q+\mu)-(E_q-\mu)\right]}
\end{eqnarray}
and
\begin{eqnarray}\nonumber\label{a_state_2nd_comb}
\Omega^{(2)}_{a}&\equiv&\langle 0|a_{\bm q}b_{\bm k - \bm p}a_{\bm p}a^{\dagger}_{\bm p' - \bm k'}b^{\dagger}_{-\bm p'}\hat{\rho}a^{\dagger}_{\bm q} |0 \rangle
\\\nonumber
&=&(2\pi)^3\delta({\bm k - \bm p + \bm p'})\langle 0|a_{\bm q}a_{\bm p}a^{\dagger}_{\bm p' - \bm k'}a_{\bm q}^{\dagger}|0\rangle 
\frac{1}{Z} {\rm e}^{\beta\left[(E_q+\mu)-(E_q-\mu)\right]}
\\
&=&(2\pi)^6\delta^3({\bm k + \bm k'})\Big[\delta^3({\bm k - \bm p + \bm p'}){\cal V}
-(2\pi)^3\delta^3({\bm q - \bm p'+ \bm k'})\delta^3({\bm p - \bm q})\Big] \nonumber \\
&& \times \frac{1}{Z} {\rm e}^{\beta\left[(E_q+\mu)-(E_q-\mu)\right]}\,.
\end{eqnarray}
Note that part of eq.~(\ref{a_state_2nd_comb}) is similar to eq.~(\ref{gstate_comb}), which does not correlate ${\bm q}$ and other momenta. Such terms actually contribute to the vacuum expectation value in the end. We hence separate the vacuum and thermal parts as $\Omega^{(2)}_{a}=\Omega^{\text{vac}}_{a}+\Omega^{\text{th}(2)}_{a}$, where 
\begin{align}
\label{Omega_a_vac}
\Omega^{\text{vac}}_{a}&=(2\pi)^6\delta^3({\bm k + \bm k'})\delta^3({\bm k - \bm p + \bm p'}){\cal V} \frac{1}{Z}{\rm e}^{\beta\left[(E_q+\mu)-(E_q-\mu)\right]}\,,
\\
\Omega^{\text{th}(2)}_{a}&=-(2\pi)^9\delta^3({\bm k + \bm k'})
\delta^3({\bm p'- \bm p + \bm k})\delta^3({\bm q - \bm p}) \frac{1}{Z} {\rm e}^{\beta\left[(E_q+\mu)-(E_q-\mu)\right]}\,.
\end{align}
For $\langle 0|b_{\bm q}\hat{M}_{\lambda\lambda'}\hat{\rho} b_{\bm q}^{\dagger}|0 \rangle$, we find 
\begin{eqnarray}\nonumber
\Omega^{\text{th}(1)}_{b}&\equiv&\langle 0|b_{\bm q}b_{\bm k - \bm p}b^{\dagger}_{- \bm p}b_{\bm k' - \bm p'}b^{\dagger}_{-\bm p'}\hat{\rho}b^{\dagger}_{\bm q} |0 \rangle
\\\nonumber
&=& \frac{1}{Z} \langle 0|b_{\bm k - \bm p}\big[(2\pi)^3\delta^3({\bm q + \bm p})-b^{\dagger}_{- \bm p}b_{\bm q}\big] \big[(2\pi)^3\delta^3({\bm k' - \bm p' - \bm q})-b^{\dagger}_{\bm q}b_{\bm k' - \bm p'} \big] b^{\dagger}_{-\bm p'}|0 \rangle
\\
&=& \frac{1}{Z} (2\pi)^9\delta^3({\bm k + \bm k'})\delta^3({\bm k - \bm p + \bm p'})\delta^3({\bm q + \bm p})\,.
\end{eqnarray}
Furthermore, 
\begin{eqnarray}\label{b_state_2nd_comb}
\Omega^{(2)}_{b}\equiv\langle 0|b_{\bm q}b_{\bm k - \bm p}a_{\bm p}a^{\dagger}_{\bm p' - \bm k'}b^{\dagger}_{-\bm p'}\hat{\rho}b^{\dagger}_{\bm q} |0 \rangle
=\Omega^{\text{vac}}_{b}+\Omega^{\text{th}(2)}_{b},
\end{eqnarray}
where 
\begin{align}
\label{Omega_b_vac}
\Omega^{\text{vac}}_{b} &= \frac{1}{Z} (2\pi)^6\delta^3({\bm k + \bm k'})\delta^3({\bm k - \bm p + \bm p'}){\cal V}\,,
\\
\Omega^{\text{th}(2)}_{b} &=
-\frac{1}{Z} (2\pi)^9\delta^3({\bm k + \bm k'})\delta^3({\bm k - \bm p + \bm p'})\delta^3({\bm q + \bm p'})\,.
\end{align}
Finally, for $\langle 0|a_{\bm q}b_{\bm q}\hat{M}_{\lambda\lambda'}\hat{\rho}b_{\bm q}^{\dagger} a_{\bm q}^{\dagger}|0 \rangle$, the first two non-vanishing and non-trivial terms are
\begin{eqnarray}
\Omega^{\text{th}(1)}_{ab} &\equiv& \langle 0|a_{\bm q}b_{\bm q}a^{\dagger}_{\bm p - \bm k}b^{\dagger}_{-\bm p}b_{\bm k' - \bm p'}a_{\bm p'}\hat{\rho} b^{\dagger}_{\bm q}a^{\dagger}_{\bm q}|0 \rangle \nonumber \\
&=& (2\pi)^{12}\delta^3({\bm k + \bm k'})\delta^3({\bm k - 2 \bm p})\delta^3({\bm q - \bm p'})\delta^3({\bm p + \bm p'})
\frac{1}{Z} {\rm e}^{-\beta(E_q-\mu)}
\end{eqnarray} 
and
\begin{eqnarray}\label{ab_state_2nd_comb}\nonumber
\Omega^{(2)}_{ab}&\equiv&\langle 0|a_{\bm q}b_{\bm q}b_{\bm k - \bm p}a_{\bm p}a^{\dagger}_{\bm p' - \bm k'}b^{\dagger}_{-\bm p'}\hat{\rho}b^{\dagger}_{\bm q} a^{\dagger}_{\bm q}|0 \rangle
\\\nonumber
&=&(2\pi)^6\delta^3({\bm k + \bm k'})\delta^3({\bm k - \bm p + \bm p'})\Big[{\cal V}^2
-{\cal V}(2\pi)^3\delta^3({\bm p - \bm q})-{\cal V}(2\pi)^3\delta({\bm q + \bm p'})
\\
&&+(2\pi)^6\delta^3({\bm q - \bm p})\delta^3({\bm q - \bm p' - \bm k'})\Big]\frac{1}{Z} {\rm e}^{-\beta(E_q-\mu)}\,,
\end{eqnarray}
which lead to $\Omega^{(2)}_{ab}=\Omega^{\text{vac}}_{ab}+\Omega^{\text{th}(2)}_{ab}$, where
\begin{align}
\label{Omega_ab_vac}
\Omega^{\text{vac}}_{ab}&=(2\pi)^6\delta^3({\bm k + \bm k'})\delta^3({\bm k - \bm p + \bm p'}){\cal V}^2\frac{1}{Z}{\rm e}^{-\beta(E_q-\mu)}\,,
\\
\Omega^{\text{th}(2)}_{ab}&=(2\pi)^{12}\delta^3({\bm k + \bm k'})\delta^3({\bm k + 2 \bm q})\delta^3({\bm k - \bm p - \bm q})\delta^3({\bm q - \bm p})\frac{1}{Z}{\rm e}^{-\beta(E_q-\mu)}
\nonumber \\
& \quad -(2\pi)^9{\cal V}\delta^3({\bm k + \bm k'})\delta^3({\bm k - \bm p + \bm p'})\Big[\delta^3({\bm p - \bm q})+\delta^3({\bm p' + \bm q})\Big] \frac{1}{Z} {\rm e}^{-\beta(E_q-\mu)}\,.
\end{align}
Further non-vanishing terms are found as  
\begin{eqnarray}
\Omega^{\text{th}(3)}_{ab} &\equiv& \langle 0|a_{\bm q}b_{\bm q}a^{\dagger}_{\bm p - \bm k}a_{\bm p}a^{\dagger}_{\bm p' - \bm k'}a_{\bm p'}\hat{\rho}b^{\dagger}_{\bm q} a^{\dagger}_{\bm q}|0 \rangle \nonumber \\
&=& (2\pi)^9{\cal V}\delta^3({\bm k + \bm k'})\delta^3({\bm k - \bm p + \bm p'})\delta^3({\bm p'- \bm q})\frac{1}{Z} {\rm e}^{-\beta(E_q-\mu)}\,,
\end{eqnarray}
and
\begin{eqnarray}\nonumber
\Omega^{\text{th}(4)}_{ab}&\equiv&\langle 0|a_{\bm q}b_{\bm q}b_{\bm k - \bm p}b^{\dagger}_{-\bm p}b_{\bm k' - \bm p'}b^{\dagger}_{-\bm p'}\hat{\rho}b^{\dagger}_{\bm q} a^{\dagger}_{\bm q}|0 \rangle
\\
&=&(2\pi)^9{\cal V}\delta^3({\bm k + \bm k'})\delta^3({\bm k - \bm p + \bm p'})\delta^3({\bm p + \bm q})\frac{1}{Z} {\rm e}^{-\beta(E_q-\mu)}\,.
\end{eqnarray}

It is clear that the combination of eqs.~(\ref{gstate_comb}), (\ref{Omega_a_vac}), (\ref{Omega_b_vac}) and (\ref{Omega_ab_vac}) yields the vacuum expectation value in eq.~(\ref{vac_comb}),
\begin{eqnarray}
\Omega_{0}+\int_{\bm q}\Omega^{\text{vac}}_a+\int_{\bm q}\Omega^{\text{vac}}_b+\frac{1}{\cal V}\int_{\bm q}\Omega^{\text{vac}}_{ab}=(2\pi)^6\delta^3({\bm k + \bm k'})\delta^3({\bm k - \bm p + \bm p'}).
\end{eqnarray}
The wave-function parts for all these terms take the same form and thus their full contribution just matches the result in vacuum.  
We can accordingly decompose 
\begin{eqnarray}\label{M_decomp}
M_{\lambda\lambda'}(\bm k,\bm k';\tau',\tau'')=M^{\text{vac}}_{\lambda\lambda'}(\bm k,\bm k';\tau',\tau'')+M^{\text{th}}_{\lambda\lambda'}(\bm k,\bm k';\tau',\tau''),
\end{eqnarray} 
where $M^{\text{vac}}_{\lambda\lambda'}(\bm k,\bm k';\tau',\tau'')$ takes the same form as eq.~(\ref{M_in_beta}). In the end, only $\Omega^{\text{th}(1)}_{a/b}$, $\Omega^{\text{th}(2)}_{a/b}$, $\Omega^{\text{th}(3,4)}_{ab}$, 
and the term linear to ${\cal V}$ in $\Omega^{\text{th}(2)}_{ab}$ contribute to the thermal expectation part, $M^{\text{th}}_{\lambda\lambda'}(\bm k,\bm k';\tau',\tau'')$, while the contributions from $\Omega^{\text{th}(1)}_{ab}$ and the remaining part of $\Omega^{\text{th}(2)}_{ab}$ vanish as will be shown later. 
We first compute the thermal contributions from $\Omega^{\text{th}(1)}_{a}$ and $\Omega^{\text{th}(2)}_{a}$, 
\begin{eqnarray}\label{Mth_a}\nonumber
&& \int_{\bm q}\langle 0|a_{\bm q}\hat{M}_{\lambda\lambda'} \hat{\rho} a^{\dagger}_{\bm q}|0\rangle_{\text{th}} \\
=&&(2\pi)^3\delta^3({\bm k + \bm k'})\int_{\bm p}\Big[u^{\dagger}({\bm p - \bm k},\tau')u(\bm p,\tau')u^{\dagger}(\bm p,\tau'')u({\bm p - \bm k},\tau'')
\chi^{(a1)}_{\lambda\lambda'}({\bm k, \bm p}) \nonumber
\\
&&+v^{\dagger}({\bm k - \bm p},\tau')u(\bm p,\tau')u^{\dagger}(\bm p,\tau'')v({\bm k - \bm p},\tau'')
\chi^{(a2)}_{\lambda\lambda'}({\bm k, \bm p})\Big]. 
\end{eqnarray}
The specific expressions for $\chi^{(a1)}_{\lambda\lambda'}({\bm k, \bm p})$ and $\chi^{(a2)}_{\lambda\lambda'}({\bm k, \bm p})$ can be derived in 
analogy to the computation for $\beta_{\lambda\lambda'}({\bm k, \bm p})$ in vacuum. By definition, 
$\chi^{(a1)}_{\lambda\lambda'}({\bm k, \bm p})$ takes the form
\begin{eqnarray}\nonumber
\chi^{(a1)}_{\lambda\lambda'}({\bm k, \bm p})&=&\frac{1}{4}(1-f_{|{\bm k - \bm p}|+\mu})f_{|{\bm k - \bm p}|-\mu}
\xi^{\dagger}_{-}({\bm p - \bm k})\epsilon^{\mu\nu}_{-\lambda}(\bm k)\big[\bar{\sigma}_{\mu}p_{\nu}+\bar{\sigma}_{\mu}(p-k)_{\nu}\big]\xi_{-}(\bm p)
\\
&&\times\xi_{-}^{\dagger}(\bm p)\epsilon^{\rho\sigma}_{-\lambda'}(-\bm k)
\big[\bar{\sigma}_{\rho}(p-k)_{\sigma}+\bar{\sigma}_{\rho}p_{\sigma}\big]\xi_{-}({\bm p - \bm k}).
\end{eqnarray}
Using the properties of $\xi_{-}$ and $\epsilon_{\lambda}^{\mu}$, we have
\begin{eqnarray}\nonumber
\chi^{(a1)}_{\lambda\lambda'}({\bm k, \bm p})&=&\frac{1}{2}(1-f_{|{\bm k - \bm p}|+\mu})f_{|{\bm k - \bm p}|-\mu}\big(\epsilon_{-\lambda}(\bm k)\cdot p\big)\big(\epsilon_{\lambda'}(\bm k)\cdot p\big)
\Bigg[\frac{2\big(\epsilon_{-\lambda}(\bm k)\cdot p\big)\big(\epsilon_{\lambda'}(\bm k)\cdot p\big)}{p|{\bm k - \bm p}|} \\
&& +\delta_{\lambda,\lambda'}\bigg(1-\frac{p^2-{\bm k\cdot \bm p}}{p|{\bm k - \bm p}|}\bigg)
-{\rm i}\epsilon_{\mu\beta\rho\alpha}
\epsilon^{\mu}_{-\lambda}(\bm k)\epsilon^{\rho}_{\lambda'}(\bm k)n^{\alpha}({\bm p - \bm k})n^{\beta}(\bm p)
\Bigg].
\end{eqnarray} 
Using the same fixed coordinate system as in the vacuum calculations, we eventually derive 
\begin{eqnarray}\nonumber
&&  \chi^{(a1)}_{\lambda\lambda'}({\bm k, \bm p}) \\
= && \frac{p^2}{4}(1-f_{|{\bm k - \bm p}|+\mu})f_{|{\bm k - \bm p}|-\mu}\sin^2\theta\big[\cos^2\phi+\lambda\lambda'\sin^2\phi+{\rm i}(\lambda'-\lambda)\cos\phi\sin\phi\big] \nonumber \\
&& \times \Bigg \{\delta_{\lambda,\lambda'}-\frac{\lambda+\lambda'}{2}\cos\theta
+\frac{1}{|{\bm k - \bm p}|}\bigg(\delta_{\lambda,\lambda'}\big(k\cos\theta-p\big)
+\frac{\lambda+\lambda'}{2}\big(p\cos\theta-k\big) \nonumber \\
&& +p\sin^2\theta\big[\cos^2\phi+\lambda\lambda'\sin^2\phi
+{\rm i}(\lambda'-\lambda)\cos\phi\sin\phi\big]
\bigg) \Bigg \}\,,
\end{eqnarray}
which yields
\begin{align}
\chi^{(a1)}_{\lambda\lambda}({\bm k, \bm p})&=\frac{p^2}{4}(1-f_{|{\bm k - \bm p}|+\mu})f_{|{\bm k - \bm p}|-\mu}\sin^2\theta \Bigg[1-\lambda\cos\theta
-(\cos\theta-\lambda)\frac{p\cos\theta-k}{|{\bm k - \bm p}|}
\Bigg]\,,
\\
\chi^{(a1)}_{\lambda,-\lambda}({\bm k, \bm p})
& =\frac{p^3}{4} (1-f_{|{\bm k - \bm p}|+\mu})f_{|{\bm k - \bm p}|-\mu}
\frac{\sin^4\theta\big(\cos 4\phi-{\rm i}\lambda\sin 4\phi\big)}{|{\bm k - \bm p}|}\,.
\end{align}
Since $|{\bm k - \bm p}|$ is independent of $\phi$, only $\chi^{(a1)}_{\lambda\lambda}({\bm k, \bm p})$ contributes to the correlation function, in analogy to the vacuum case. 
We hence have 
\begin{eqnarray}\label{chia1_same _helicity}
\chi^{(a1)}_{\lambda\lambda'}({\bm k, \bm p})&=&\delta_{\lambda\lambda'}\frac{p^2}{4}(1-f_{|{\bm k - \bm p}|+\mu})f_{|{\bm k - \bm p}|-\mu}\sin^2\theta \nonumber \\
&& \times \Bigg[1-\lambda\cos\theta
-(\cos\theta-\lambda)\frac{p\cos\theta-k}{\sqrt{p^2-2pk\cos\theta+k^2}}
\Bigg].
\end{eqnarray}
$\chi^{(a2)}_{\lambda\lambda'}({\bm k, \bm p})$ is obtained as 
\begin{eqnarray}
\chi^{(a2)}_{\lambda\lambda'}({\bm k, \bm p})&=&-(1-f_{p+\mu})f_{p-\mu}\beta_{\lambda\lambda'}({\bm k, \bm p}).
\end{eqnarray}

Next, we evaluate the thermal contribution from $\Omega^{\text{th}}_{b}$, which gives
\begin{eqnarray}\label{Mth_b}\nonumber
\int_{\bm q}\langle 0|b_{\bm q}\hat{M}_{\lambda\lambda'}b^{\dagger}_{\bm q}|0\rangle_{\text{th}}
&=&(2\pi)^3\delta^3({\bm k + \bm k'})\int_{\bm p}\Big[v^{\dagger}({\bm k - \bm p},\tau')v(-\bm p,\tau')v^{\dagger}(-\bm p,\tau'')v({\bm k - \bm p},\tau'')\chi^{(b1)}_{\lambda\lambda'}({\bm k, \bm p})
\\
&&+v({\bm k - \bm p},\tau')u(\bm p,\tau')u^{\dagger}(\bm p,\tau'')v^{\dagger}({\bm k - \bm p},\tau'')\chi^{(b2)}_{\lambda\lambda'}({\bm k, \bm p})\Big].
\end{eqnarray}
$\chi^{(b1)}_{\lambda\lambda'}({\bm k, \bm p})$ here is defined as
\begin{eqnarray}\nonumber
\chi^{(b1)}_{\lambda\lambda'}({\bm k, \bm p})&=&\frac{1}{4}(1-f_{p-\mu})f_{p+\mu}
\xi^{\dagger}_{-}({\bm k - \bm p})\epsilon^{\mu\nu}_{-\lambda}(\bm k)\big[\bar{\sigma}_{\mu}p_{\nu}+\bar{\sigma}_{\mu}(p-k)_{\nu}\big]\xi_{-}(-\bm p)
\\
&&\times\xi_{-}^{\dagger}(-\bm p)\epsilon^{\rho\sigma}_{-\lambda'}(-\bm k)
\big[\bar{\sigma}_{\rho}(p-k)_{\sigma}+\bar{\sigma}_{\rho}p_{\sigma}\big]\xi_{-}({\bm k - \bm p}).
\end{eqnarray}
Following a computation similar to that for $\chi_{\lambda\lambda'}^{(a1)}({\bm k, \bm p})$, we find
\begin{eqnarray}\nonumber
\chi^{(b1)}_{\lambda\lambda'}({\bm k, \bm p})
&=&\frac{1}{2}(1-f_{p-\mu})f_{p+\mu}\big(\epsilon_{-\lambda}(\bm k)\cdot p\big)\big(\epsilon_{\lambda'}(\bm k)\cdot p\big)
\Bigg[\frac{2\big(\epsilon_{-\lambda}(\bm k)\cdot p\big)\big(\epsilon_{\lambda'}(\bm k)\cdot p\big)}{p|{\bm k - \bm p}|} \\
&& +\delta_{\lambda,\lambda'}\bigg(1-\frac{p^2-{\bm k\cdot \bm p}}{p|{\bm k - \bm p}|}\bigg)
-{\rm i}\epsilon_{\mu\beta\rho\alpha}
\epsilon^{\mu}_{-\lambda}(\bm k)\epsilon^{\rho}_{\lambda'}(\bm k)n^{\alpha}({\bm k - \bm p})n^{\beta}(-\bm p)
\Bigg]\,.
\end{eqnarray}
Employing again the same fixed coordinate system as in the vacuum calculation, we eventually derive 
\begin{eqnarray}\nonumber
&& \chi^{(b1)}_{\lambda\lambda'}({\bm k, \bm p}) \\
=&&\frac{p^2}{4}(1-f_{p-\mu})f_{p+\mu}\sin^2\theta\big[\cos^2\phi+\lambda\lambda'\sin^2\phi+{\rm i}(\lambda'-\lambda)\cos\phi\sin\phi\big] \nonumber \\
&& \times \Bigg[\delta_{\lambda,\lambda'}+\frac{\lambda+\lambda'}{2}\cos\theta
+\frac{1}{|{\bm k - \bm p}|}\bigg(\delta_{\lambda,\lambda'}\big(k\cos\theta-p\big)
+\frac{\lambda+\lambda'}{2}\big(k-p\cos\theta\big) \nonumber \\
&& +p\sin^2\theta\big[\cos^2\phi+\lambda\lambda'\sin^2\phi +{\rm i}(\lambda'-\lambda)\cos\phi\sin\phi\big]
\bigg)\Bigg],
\end{eqnarray}
which yields
\begin{align}
\label{chib_same _helicity}
\chi^{(b1)}_{\lambda\lambda}({\bm k, \bm p})&=\frac{p^2}{4}(1-f_{p-\mu})f_{p+\mu}\sin^2\theta \Bigg[1+\lambda\cos\theta
-(\cos\theta+\lambda)\frac{p\cos\theta-k}{|{\bm k - \bm p}|}
\Bigg]\,,
\\
\chi^{(b1)}_{\lambda,-\lambda}({\bm k, \bm p})
&=\frac{p^3}{4}(1-f_{p-\mu})f_{p+\mu}
\frac{\sin^4\theta\big(\cos 4\phi-{\rm i}\lambda\sin 4\phi\big)}{|{\bm k - \bm p}|}\,.
\end{align}
Again, only $\chi^{(b1)}_{\lambda\lambda}({\bm k, \bm p})$ contributes to the correlation function, leading to
\begin{eqnarray}\label{chib1_same _helicity}
\chi^{(b1)}_{\lambda\lambda'}({\bm k, \bm p})&=& \delta_{\lambda\lambda'} \frac{p^2}{4}(1-f_{p-\mu})f_{p+\mu}\sin^2\theta \Bigg[1+\lambda\cos\theta
-(\cos\theta+\lambda)\frac{p\cos\theta-k}{|{\bm k - \bm p}|}
\Bigg]\,.
\end{eqnarray}
Furthermore, $\chi_{\lambda\lambda'}^{(b2)}({\bm k, \bm p})$ is obtained as
\begin{eqnarray}
\chi^{(b2)}_{\lambda\lambda'}({\bm k, \bm p})&=&-(1-f_{|{\bm k - \bm p}|-\mu})f_{|{\bm k - \bm p}|+\mu}\beta_{\lambda\lambda'}({\bm k, \bm p}).
\end{eqnarray}

Finally, the contribution from $\Omega^{\text{th}(1-4)}_{ab}$ reads
\begin{eqnarray}\nonumber
&& \frac{1}{\cal V}\int_{\bm q}\langle 0|a_{\bm q}b_{\bm q}\hat{M}_{\lambda\lambda'}b_{\bm q}^{\dagger}a^{\dagger}_{\bm q}|0\rangle_{\text{th}} \\
=&&(2\pi)^3\delta^3({\bm k + \bm k'})\int_{\bm p}\Big[u^{\dagger}({\bm p - \bm k},\tau')v(-\bm p,\tau')v^{\dagger}(-\bm p,\tau'')u({\bm p - \bm k},\tau'')\chi^{(ab1)}_{\lambda\lambda'}({\bm k, \bm p})
\nonumber \\
&&+v^{\dagger}({\bm k - \bm p},\tau')u(\bm p,\tau')u^{\dagger}(\bm p,\tau'')v({\bm k - \bm p},\tau'')\chi^{(ab2)}_{\lambda\lambda'}({\bm k, \bm p})\Big],
\end{eqnarray}
where
\begin{align}
\chi^{(ab1)}_{\lambda\lambda'}({\bm k, \bm p}) &=
\frac{f_{|{\bm k - \bm p}|-\mu}f_{|{\bm k - \bm p}|+\mu}\chi^{(a1)}_{\lambda\lambda'}({\bm k, \bm p})}{(1-f_{|{\bm k - \bm p}|+\mu})f_{|{\bm k - \bm p}|-\mu}}+\frac{f_{p-\mu}f_{p+\mu}\chi^{(b1)}_{\lambda\lambda'}({\bm k, \bm p})}{(1-f_{p-\mu})f_{p+\mu}}\,,
\\
\chi^{(ab2)}_{\lambda\lambda'}({\bm k, \bm p}) &=-\big(f_{|{\bm k - \bm p}|-\mu}f_{|{\bm k - \bm p}|+\mu}+f_{p-\mu}f_{p+\mu}\big)\beta_{\lambda\lambda'}({\bm k, \bm p})\,.
\end{align}
Here $\chi^{(ab1)}_{\lambda\lambda'}({\bm k, \bm p})$ comes from $\Omega^{\text{th}(3,4)}_{ab}$, while $\chi^{(ab2)}_{\lambda\lambda'}({\bm k, \bm p})$ originates from the terms linear in ${\cal V}$ in $\Omega^{\text{th}(2)}_{ab}$.

In the conformal coordinates, the wave functions reduce to plane waves such that $u(\bm p,\tau)={\rm e}^{-{\rm i}p\tau}$ and $v(\bm p,\tau)={\rm e}^{{\rm i}p\tau}$.
It is then easy to recognize that $\chi^{(ab1)}_{\lambda\lambda'}({\bm k, \bm p})$ and $\chi^{(ab2)}_{\lambda\lambda'}({\bm k, \bm p})$ take similar forms as $\chi^{(a,b1)}_{\lambda\lambda'}({\bm k, \bm p})$ and $\chi^{(a,b2)}_{\lambda\lambda'}({\bm k, \bm p})$, respectively. 
Furthermore, one finds that the contributions from $\Omega^{\text{th}(1)}_{ab}$ and the first term of $\Omega^{\text{th}(2)}_{ab}$ vanish due to their on-shell conditions, $\bm p=\bm k/2$ and $\bm p=-\bm k/2$, respectively. 

Adding all non-vanishing terms and replacing the wave functions with the abovementioned plane wave solutions, we obtain 
eqs.\,(\ref{eq:hcor_th})--(\ref{eq:Gamma_bar_0}).

\section{Derivation of eqs.\,(\ref{eq:eq12}) and (\ref{eq:eq21})}
\label{sec:tau_integrals}
We here present the details of the calculations related to the $\tau'$ integrals. 
Substituting eqs.\,(\ref{eq:scale_factor}) and (\ref{eq:G_solution}) into eq.\,(\ref{eq:eq1}) and rearanging some terms, we obtain 
\begin{align}
\bar{\mathcal{I}}(\tau,\tau_{\rm i},b) = \frac{1}{2{\rm i}} \frac{\tau_{\rm i}^2}{a_{\rm i}^2 k \tau} 
\left[
{\rm e}^{{\rm i} \tau k} \int_{\tau_{\rm i}}^{\tau} \frac{{\rm d} \tau'}{\tau'} {\rm e}^{{\rm i} \tau' (b - k)} 
- {\rm e}^{-{\rm i} \tau k} \int_{\tau_{\rm i}}^{\tau} \frac{{\rm d} \tau'}{\tau'} {\rm e}^{{\rm i} \tau' (b + k)}
\right], 
\label{eq:eq6.1}
\end{align}
which will be the starting point for the following considerations.

\subsection{$b = b_-$ case}
To derive eq.\,(\ref{eq:eq12}), we first expand $|\bm{k} - \bm{p}|$, which gives 
\begin{eqnarray}
|\bm{k} - \bm{p}| \simeq p \left(
1 - \frac{k}{p} \cos \theta + \frac{1}{2} \frac{k^2}{p^2} \sin^2 \theta 
\right),  
\label{eq:eq7}
\end{eqnarray}
where $\theta$ is the angle between $\bm{k}$ and $\bm{p}$. Thus
\begin{eqnarray}
{\rm e}^{{\rm i} \tau' (b_{-} \mp k)} \simeq  1 - {\rm i}\tau' k(\cos \theta \pm 1) + \frac{\rm i}{2} \tau' \frac{k^2}{p} \sin^2 \theta - \frac{1}{2} \tau'^2  k^2 (\cos \theta \pm 1)^2\,.
\label{eq:eq8}
\end{eqnarray}
Hence, we have
\begin{eqnarray}
\int_{\tau_{\rm i}}^{\tau} \frac{{\rm d} \tau'}{\tau'}  {\rm e}^{{\rm i} \tau' (b_{-} \mp k)} &\simeq&  
\log \Bigl(\frac{\tau}{\tau_{\rm i}} \Bigr) -{\rm i}k(\tau - \tau_{\rm i}) \Bigl( \cos \theta \pm 1 - \frac{1}{2} \frac{k}{p} \sin^2 \theta \Bigr) \nonumber \\
&&-\frac{1}{4}(\tau^2 - \tau^2_0) k^2 (\cos \theta \pm 1)^2\,.
\label{eq:eq9}
\end{eqnarray}
Note that because the LO term cancels in eq.\,(\ref{eq:eq6.1}), terms up to second order in $k$ have to be kept up to this 
point of the calculation. 
Substituting the above results into eq.\,(\ref{eq:eq6.1}) and expanding the factors ${\rm e}^{{\rm i} \tau k}$ and ${\rm e}^{-{\rm i} \tau k}$ 
in $k$, we obtain 
\begin{eqnarray}
\bar{\mathcal{I}}(\tau,\tau_{\rm i},b_-) \simeq 
\frac{\tau_{\rm i}^2}{a_{\rm i}^2} \left[
\log \Bigl( \frac{\tau}{\tau_{\rm i}} \Bigr) - \frac{\tau - \tau_{\rm i}}{\tau} 
- \frac{\rm i}{2} k \frac{(\tau - \tau_{\rm i})^2}{\tau} 
\right] + \mathcal{O}(k^2)\,. 
\label{eq:eq11}
\end{eqnarray}
As the zeroth-order term in $k$ is real while the first-order term is imaginary, 
the latter becomes a second-order contribution to $|\bar{\mathcal{I}}(\tau,\tau_{\rm i},b_-)|^2$ and can hence be neglected. 
Therefore, we have 
\begin{eqnarray}
|\bar{\mathcal{I}}(\tau,\tau_{\rm i},b_-)|^2 \simeq 
\frac{\tau_{\rm i}^4}{a_{\rm i}^4} \left[
\log \Bigl( \frac{\tau}{\tau_{\rm i}} \Bigr) - \frac{\tau - \tau_{\rm i}}{\tau} 
\right]^2 
+ \mathcal{O}(k^2)\,. 
\label{eq:eq12_app}
\end{eqnarray}

\subsection{$b = b_+$ case}
Using eq.\,(\ref{eq:eq7}), we have 
\begin{eqnarray}
{\rm e}^{{\rm i} \tau' (b_+ \mp k)} \simeq {\rm e}^{2 {\rm i} \tau' p} \left[ 
1 - {\rm i} \tau' k(\cos \theta \pm 1) + \frac{\rm i}{2} \tau' \frac{k^2}{p} \sin^2 \theta - \frac{1}{2} \tau'^2  k^2 (\cos \theta \pm 1)^2
\right], 
\label{eq:eq13}
\end{eqnarray}
which leads to 
\begin{eqnarray}
&& \int_{\tau_{\rm i}}^{\tau} \frac{{\rm d}\tau'}{\tau'}  {\rm e}^{{\rm i} \tau' (b_+ \mp k)} \nonumber \\ 
\simeq &&  
\mathrm{Ei}(2{\rm i}\tau p) - \mathrm{Ei}(2{\rm i}\tau_{\rm i} p)  + \frac{k}{p} {\rm e}^{{\rm i}(\tau + \tau_{\rm i})p} \Biggl\{
-{\rm i}\sin [(\tau - \tau_{\rm i}) p] (\cos \theta \pm 1) \nonumber \\
&& +\frac{\rm i}{2} \frac{k}{p} \sin [(\tau - \tau_{\rm i}) p] \sin^2 \theta - \frac{1}{4} \frac{k}{p} 
\Bigl( ({\rm i} + (\tau + \tau_{\rm i})p) \sin [(\tau - \tau_{\rm i}) p]  \nonumber \\
&& - {\rm i}(\tau - \tau_{\rm i}) p \cos [(\tau - \tau_{\rm i}) p]
\Bigr) (\cos \theta \pm 1)^2
\Biggr\}\,,
\label{eq:eq14}
\end{eqnarray}
where $Ei(x)$ is the exponential integral function. 
Substituting the above results into eq.\,(\ref{eq:eq6.1}) and expanding the factors ${\rm e}^{{\rm i} \tau k}$ and ${\rm e}^{-{\rm i}\tau k}$ 
in $k$, we obtain 
\begin{eqnarray}
&& \bar{\mathcal{I}}(\tau,\tau_{\rm i},b_+) \nonumber \\
\simeq && 
\frac{\tau_{\rm i}^2}{a_{\rm i}^2} \left[
\mathrm{Ei}(2{\rm i}\tau p) - \mathrm{Ei}(2{\rm i}\tau_{\rm i} p) 
- \frac{1}{\tau p} {\rm e}^{{\rm i}(\tau + \tau_{\rm i})p} \sin [(\tau - \tau_{\rm i}) p]  (1 + {\rm i} \tau k \cos \theta )
\right] \nonumber \\
&&+ \mathcal{O}(k^2). 
\label{eq:eq16}
\end{eqnarray}
For the next step, we make use of
\begin{eqnarray}
\mathrm{Ei}({\rm i}x) = \mathrm{ci}(x) + {\rm i} \mathrm{Si}(x), 
\label{eq:eq17}
\end{eqnarray}
for $x$ real and positive, where $\mathrm{ci}(x)$ and $\mathrm{Si}(x)$ are the cosine and sine integral 
functions defined in eqs.~(\ref{eq:ci}) and (\ref{eq:Si}), respectively.
Via eq.\,(\ref{eq:eq17}), we can now divide $\bar{\mathcal{I}}(\tau,\tau_{\rm i},b)$ into its real and imaginary parts, 
and hence obtain 
\begin{eqnarray}
&& |\bar{\mathcal{I}}(\tau,\tau_{\rm i},b_+)|^2 \nonumber \\ 
\simeq &&  
\frac{\tau_{\rm i}^4}{a_{\rm i}^4} \Biggl \{ \Bigl(
\mathrm{ci}(2\tau p) - \mathrm{ci}(2\tau_{\rm i} p) - \frac{1}{\tau p} \sin [(\tau - \tau_{\rm i}) p]  \cos [(\tau + \tau_{\rm i}) p] \Bigr)^2 \nonumber \\
&&+\Bigl(
\mathrm{Si}(2\tau p) - \mathrm{Si}(2\tau_{\rm i} p) - \frac{1}{\tau p} \sin [(\tau - \tau_{\rm i}) p] \sin [(\tau + \tau_{\rm i}) p] \Bigr)^2  \nonumber \\
&& + 2\frac{k}{ p} \cos \theta  \sin [(\tau - \tau_{\rm i}) p] 
\Bigr(
\sin [(\tau + \tau_{\rm i}) p] 
\bigl[
\mathrm{ci}(2\tau p) - \mathrm{ci}(2\tau_{\rm i} p) \bigr]  \nonumber \\
&&- \cos [(\tau + \tau_{\rm i}) p] 
\bigl[
\mathrm{Si}(2\tau p) - \mathrm{Si}(2\tau_{\rm i} p) \bigr] \Bigr)
\Biggr \} + \mathcal{O}(k^2)\,. 
\label{eq:eq21_app}
\end{eqnarray}

\section{Derivation of eqs.\,(\ref{eq:eq26})--(\ref{eq:eq34.6})}
\label{sec:angular_integrals}
We in this appendix provide the details of the angular integrals of $\phi$ and $\theta$ of $\bm{p}$ in eq.\,(\ref{eq:hcor_th}). 
As the $\phi$ integral is trivial, we only need to discuss the $\theta$ integral explicitly.

\subsection{$\Gamma_{\pm}$ terms}
Before considering the $\theta$ integrals of these terms, we first need to expand eq.\,(\ref{eq:Gamma_bar_pm}) 
in small $k$. Making use of eq.\,(\ref{eq:eq7}), the result reads
\begin{eqnarray}
\tilde{\Gamma}_{\pm}(\bm{p},\theta;\bm{k},\lambda) \simeq \sin^4 \theta 
\left[ 1 + \frac{k}{ p} (\cos \theta \mp \lambda) \right] + \mathcal{O}(k^2).  
\label{eq:eq22}
\end{eqnarray}
Furthermore, for the $\Gamma_{+}$ term, we also need to expand $f_{|\bm{k} - \bm{p}| - \bar \mu}$, which gives 
\begin{eqnarray}
f_{|\bm{k} - \bm{p}| - \bar \mu} \simeq 
f_{p - \bar \mu} \left[
1 + \frac{k}{\bar T} \cos \theta (1 - f_{p - \bar \mu})
\right]
 + \mathcal{O}(k^2).   
\label{eq:eq23}
\end{eqnarray}

Next, substituting the results of eqs.\,(\ref{eq:eq12}), (\ref{eq:eq22}) and (\ref{eq:eq23}) into eq.\,(\ref{eq:hcor_th}), 
we obtain for the $\Gamma_{+}$ term 
\begin{eqnarray}
&& \langle h_{\lambda}(\bm{k}, \tau) h_{\lambda'}(\bm{k}', \tau) \rangle_{\mathrm{th},\,\Gamma_{+}} \nonumber \\
\simeq && 
2 \pi \delta_{\lambda,\lambda'}\frac{\delta(\bm{k} + \bm{k}')}{m_{\rm P}^4} \frac{\tau_{\rm i}^4}{a_{\rm i}^4} 
\left[ \log \Bigl( \frac{\tau}{\tau_{\rm i}} \Bigr) - \frac{\tau - \tau_{\rm i}}{\tau} 
\right]^2 \int_0^{\infty} {\rm d}p \ p^4 f_{p - \bar \mu} \nonumber \\
&& \times \int_0^{\pi} {\rm d} \theta \sin^5 \theta 
\left[
1 + \frac{k}{ p} (\cos \theta - \lambda)
\right]
\left[
1 + \frac{k}{\bar T} \cos \theta (1 - f_{p - \bar \mu})
\right]
 + \mathcal{O}(k^2)\,,   
\label{eq:eq24}
\end{eqnarray}
Using 
\begin{eqnarray}
\int_0^{\pi} {\rm d} \theta \sin^5 \theta & = & \frac{16}{15}\,, \\
\int_0^{\pi} {\rm d} \theta \sin^5 \theta \cos \theta & = & 0\,, 
\label{eq:eq25}
\end{eqnarray}
we have
\begin{eqnarray}
&& \langle h_{\lambda}(\bm{k}, \tau) h_{\lambda'}(\bm{k}', \tau) \rangle_{\mathrm{th},\,\Gamma_{+}} \nonumber \\
\simeq &&
\frac{32 \pi}{15} \delta_{\lambda,\lambda'}\frac{\delta(\bm{k} + \bm{k}')}{m_{\rm P}^4} \frac{\tau_{\rm i}^4}{a_{\rm i}^4} 
\left[
\log \Bigl( \frac{\tau}{\tau_{\rm i}} \Bigr) - \frac{\tau - \tau_{\rm i}}{\tau} 
\right]^2 \int_0^{\infty} {\rm d}p \ p^4 
\left(1 - \lambda \frac{k}{ p} \right)
f_{p - \bar \mu}  \nonumber \\
&& + \mathcal{O}(k^2).   
\label{eq:eq26_app}
\end{eqnarray}
For the $\Gamma_{-}$ term, a similar calculation leads to 
\begin{eqnarray}
&& \langle h_{\lambda}(\bm{k}, \tau) h_{\lambda'}(\bm{k}', \tau) \rangle_{\mathrm{th},\,\Gamma_{-}} \nonumber \\
\simeq &&
\frac{32 \pi}{15} \delta_{\lambda,\lambda'}\frac{\delta(\bm{k} + \bm{k}')}{m_{\rm P}^4} \frac{\tau_{\rm i}^4}{a_{\rm i}^4} 
\left[
\log \Bigl( \frac{\tau}{\tau_{\rm i}} \Bigr) - \frac{\tau - \tau_{\rm i}}{\tau} 
\right]^2 \int_0^{\infty} {\rm d}p \ p^4 
\left(1 + \lambda \frac{k}{ p} \right)
f_{p + \bar \mu}  \nonumber \\
&& + \mathcal{O}(k^2).   
\label{eq:eq27_app}
\end{eqnarray}

\subsection{$\Gamma_{0}$ term}
For handling the $\theta$ integral of this term, we first need to expand $\tilde{\Gamma}_{0}$ in eq.\,(\ref{eq:Gamma_bar_0})
and  $f_{|\bm{k} - \bm{p}| + \bar \mu}$ in small $k$,
\begin{gather}
\tilde{\Gamma}_{0}(\bm{p},\theta;\bm{k},\lambda) \simeq \sin^2 \theta \Bigl[(1 - \lambda \cos \theta)^2 
- \frac{k}{ p} \sin^2 \theta (\cos \theta - \lambda) \Bigr]
+ \mathcal{O}(k^2),  
\label{eq:eq28}
\\
f_{|\bm{k} - \bm{p}| + \bar \mu} \simeq 
f_{p + \bar \mu} \Bigl[
1 + \frac{k}{\bar T} \cos \theta (1 - f_{p + \bar \mu})
\Bigr]
 + \mathcal{O}(k^2).  
\label{eq:eq29}
\end{gather}
As the analytic structure of this term is very involved, we will consider the LO ($k^0$) and NLO ($k^1$) terms 
separately.

\subsubsection{Leading-order contribution}
Substituting the LO parts of eqs.\,(\ref{eq:eq21}), (\ref{eq:eq28}) and (\ref{eq:eq29}) into eq.\,(\ref{eq:hcor_th}), 
using
\begin{eqnarray}
\int_0^{\pi} {\rm d} \theta \sin^3 \theta  (1 - \lambda \cos \theta)^2 = \frac{8}{5}\,, 
\label{eq:eq31}
\end{eqnarray}
and noting that the $\lambda$-dependent term vanishes, we obtain
\begin{eqnarray}
&& \langle h_{\lambda}(\bm{k}, \tau) h_{\lambda'}(\bm{k}', \tau) \rangle_{\mathrm{th},\,\Gamma_{0},\,\mathrm{LO}} \nonumber \\
\simeq &&
\frac{16 \pi}{5} \delta_{\lambda,\lambda'}\frac{\delta(\bm{k} + \bm{k}')}{m_{\rm P}^4} \frac{\tau_{\rm i}^4}{a_{\rm i}^4} 
\int_0^{\infty} {\rm d}p \ p^4 (f_{p - \bar \mu} + f_{p + \bar \mu}) \nonumber \\
&& \times 
\Biggl \{ \Bigl(
\mathrm{ci}(2\tau p) - \mathrm{ci}(2\tau_{\rm i} p) - \frac{1}{\tau p} \sin [(\tau - \tau_{\rm i}) p]  \cos [(\tau + \tau_{\rm i}) p] \Bigr)^2 \nonumber \\
&&+\Bigl(
\mathrm{Si}(2\tau p) - \mathrm{Si}(2\tau_{\rm i} p) - \frac{1}{\tau p} \sin [(\tau - \tau_{\rm i}) p] \sin [(\tau + \tau_{\rm i}) p] \Bigr)^2 \Biggr \}\,.
\label{eq:eq32_app}
\end{eqnarray}

\subsubsection{Next-to-leading-order contribution}
Collecting all terms linear in $k$ from eqs.\,(\ref{eq:eq21}), (\ref{eq:eq28}) and (\ref{eq:eq29}), 
and performing the $\theta$ integrals
\begin{eqnarray}
\int_0^{\pi} {\rm d} \theta \sin^3 \theta \cos \theta  (1 - \lambda \cos \theta)^2 & = & - \frac{8}{15} \lambda\,, \\
\int_0^{\pi} {\rm d} \theta \sin^5 \theta (\cos \theta  - \lambda ) & = & - \frac{16}{15} \lambda\,, 
\label{eq:eq34.5}
\end{eqnarray}
one gets
\begin{eqnarray}
&& \langle h_{\lambda}(\bm{k}, \tau) h_{\lambda'}(\bm{k}', \tau) \rangle_{\mathrm{th},\,\Gamma_{0},\,\mathrm{NLO}} \nonumber \\
\simeq &&
-\frac{16 \pi}{15} \delta_{\lambda,\lambda'}\frac{\delta(\bm{k} + \bm{k}')}{m_{\rm P}^4} \frac{\tau_{\rm i}^4}{a_{\rm i}^4} \lambda k 
\int_0^{\infty} {\rm d}p \ p^4  \nonumber \\
&&\times \Biggl \{ \frac{2}{p} (f_{p - \bar \mu} + f_{p + \bar \mu}) \sin [(\tau - \tau_{\rm i}) p] 
\Bigr(
\sin [(\tau + \tau_{\rm i}) p] 
\bigl[
\mathrm{ci}(2\tau p) - \mathrm{ci}(2\tau_{\rm i} p) \bigr]  \nonumber \\
&&- \cos [(\tau + \tau_{\rm i}) p] 
\bigl[
\mathrm{Si}(2\tau p) - \mathrm{Si}(2\tau_{\rm i} p) \bigr] \Bigr) \nonumber \\
&& + \left[- \frac{2}{p} (f_{p - \bar \mu} + f_{p + \bar \mu}) + \frac{1}{T} f_{p + \bar \mu}  (1 - f_{p + \bar \mu}) \right]
\nonumber \\ 
&& \times \biggl[ \Bigl(
\mathrm{ci}(2\tau p) - \mathrm{ci}(2\tau_{\rm i} p) - \frac{1}{\tau p} \sin [(\tau - \tau_{\rm i}) p]  \cos [(\tau + \tau_{\rm i}) p] \Bigr)^2 \nonumber \\
&&+\Bigl(
\mathrm{Si}(2\tau p) - \mathrm{Si}(2\tau_{\rm i} p) - \frac{1}{\tau p} \sin [(\tau - \tau_{\rm i}) p] \sin [(\tau + \tau_{\rm i}) p] \Bigr)^2 \biggr] \Biggr \}\,.
\label{eq:eq34.6_app}
\end{eqnarray}

\section{Derivation of eqs.\,(\ref{eq:eq52.1}) and (\ref{eq:eq67})}
\label{sec:p_integrals}
The remaining $p$ integrals in eqs.\,(\ref{eq:eq26}), (\ref{eq:eq32}) and (\ref{eq:eq34.6}) are discussed in this appendix. 

\subsection{$\Gamma_{\pm}$ terms}
Starting from eq.\,(\ref{eq:eq26}), we need to evaluate the following integral
\begin{eqnarray}
I_1(k,\pm \bar \mu, \bar T) = \int_0^{\infty} {\rm d}p \ p^4 
\left(
1 \pm \lambda \frac{k}{ p}
\right)
f_{p \pm \bar \mu}\,.
\label{eq:I1}
\end{eqnarray}
Changing the integration variable from $p$ to $y = p/\bar T$ and making use of the following general integration formula
\begin{eqnarray}
\int_0^{\infty} {\rm d}y y^n 
\frac{1}{{\rm e}^{y \pm \bar \mu/\bar T} + 1} = - n! \mathrm{Li}_{n+1} \bigl(-{\rm e}^{\mp \bar \mu/\bar T}\bigr), 
\label{eq:eq48}
\end{eqnarray}
where $\mathrm{Li}_{s}(z) = \sum_{k=1}^{\infty} z^k/k^s$ is the polylogarithm function, we obtain
\begin{eqnarray}
I_1(k,\pm \bar \mu,\bar T) = -24 \bar T^5 \Bigl[\mathrm{Li}_{5} \bigl(-{\rm e}^{\mp \bar \mu/\bar T}\bigr)  \pm \frac{\lambda}{4} \frac{k}{\bar T} \mathrm{Li}_{4} \bigl(-{\rm e}^{\mp \bar \mu/\bar T}\bigr)  \Bigr],
\label{eq:eq50}
\end{eqnarray}
which leads to eq.\,(\ref{eq:eq52.1}).

\subsection{$\Gamma_{0}$ term}
As we discussed in the main text, it is possible to recast the functions $\mathrm{Si}(x)$ and $\mathrm{ci}(x)$ into simpler expressions for the temperature and time regions considered in this work. By making use of eqs.\,(\ref{eq:eq63.1}) and (\ref{eq:eq63}), the parts of the integrands involving $\mathrm{Si}(x)$ and $\mathrm{ci}(x)$ can be simplified as 
\begin{eqnarray}
&&\Bigl(
\mathrm{ci}(2\tau p) - \mathrm{ci}(2\tau_{\rm i} p) - \frac{1}{\tau p} \sin [(\tau - \tau_{\rm i}) p]  \cos [(\tau + \tau_{\rm i}) p] \Bigr)^2 \nonumber \\
&&+\Bigl(
\mathrm{Si}(2\tau p) - \mathrm{Si}(2\tau_{\rm i} p) - \frac{1}{\tau p} \sin [(\tau - \tau_{\rm i}) p] \sin [(\tau + \tau_{\rm i}) p] \Bigr)^2 \nonumber \\
\simeq && \frac{(\tau - \tau_{\rm i})^2}{4 \tau_{\rm i}^2 \tau^2  p^2} + \mathcal{O}\Bigl(\frac{1}{(\tau p)^3}\Bigr)\,, 
\label{eq:eq65}
\end{eqnarray}
and
\begin{eqnarray}
&& \sin [(\tau + \tau_{\rm i}) p] 
\bigl[
\mathrm{ci}(2\tau p) - \mathrm{ci}(2\tau_{\rm i} p) \bigr] 
- \cos [(\tau + \tau_{\rm i}) p] \bigl[
\mathrm{Si}(2\tau p) - \mathrm{Si}(2\tau_{\rm i} p) \bigr] \nonumber \\
= && -\frac{\tau - \tau_{\rm i}}{2 \tau_{\rm i} \tau  p} \cos [(\tau - \tau_{\rm i}) p] + \mathcal{O}\Bigl(\frac{1}{(\tau p)^2}\Bigr)\,. 
\label{eq:eq66}
\end{eqnarray}
Using eq.\,(\ref{eq:eq65}), the LO $\Gamma_{0}$ term of eq.\,(\ref{eq:eq32}) becomes 
\begin{eqnarray}
\langle h_{\lambda}(\bm{k}, \tau) h_{\lambda'}(\bm{k}', \tau) \rangle_{\mathrm{th},\,\Gamma_{0},\,\mathrm{LO}} & \simeq & 
\frac{4 \pi}{5} \delta_{\lambda,\lambda'}\frac{\delta(\bm{k} + \bm{k}')}{m_{\rm P}^4} \frac{\tau_{\rm i}^4}{a_{\rm i}^4} \frac{(\tau - \tau_{\rm i})^2}{\tau^2 \tau_{\rm i}^2} 
\int_0^{\infty} {\rm d}p \ p^2 (f_{p - \bar \mu} + f_{p + \bar \mu}) \nonumber \\
& = & \,\, -\frac{8 \pi}{5} \delta_{\lambda,\lambda'}\frac{\delta(\bm{k} + \bm{k}')}{m_{\rm P}^4} \frac{\tau_{\rm i}^4}{a_{\rm i}^4} \frac{(\tau - \tau_{\rm i})^2}{\tau^2 \tau_{\rm i}^2} \bar T^3 \nonumber \\
&& \times \Bigl[\mathrm{Li}_{3} \bigl(-{\rm e}^{\bar \mu/\bar T}\bigr) + \mathrm{Li}_{3} \bigl(-{\rm e}^{-\bar \mu/\bar T}\bigr) \Bigr]\,, 
\label{eq:eq67_app}
\end{eqnarray}
which corresponds to the LO term in eq.\,(\ref{eq:eq67}). 
Moreover, employing both eqs.\,(\ref{eq:eq65}) and (\ref{eq:eq66}), the NLO term can be simplified as 
\begin{eqnarray}
\langle h_{\lambda}(\bm{k}, \tau) h_{\lambda'}(\bm{k}', \tau) \rangle_{\mathrm{th},\,\Gamma_{0},\,\mathrm{NLO}} &\simeq &
\frac{4 \pi}{15} \delta_{\lambda,\lambda'}\frac{\delta(\bm{k} + \bm{k}')}{m_{\rm P}^4} \frac{\tau_{\rm i}^4}{a_{\rm i}^4}\frac{\tau - \tau_{\rm i}}{\tau \tau_{\rm i}} \lambda k 
\int_0^{\infty} {\rm d}p \ p^2  \nonumber \\
&&\times \Biggl \{ 2 (f_{p - \bar \mu} + f_{p + \bar \mu}) \sin [2(\tau - \tau_{\rm i}) p] 
+ \frac{\tau - \tau_{\rm i}}{\tau_{\rm i} \tau} \nonumber \\
&& \times \left[ \frac{2}{p} (f_{p - \bar \mu} + f_{p + \bar \mu})  
 - \frac{1}{\bar T} f_{p + \bar \mu}  (1 - f_{p + \bar \mu}) \right] \Biggr \}\,.
\label{eq:eq68}
\end{eqnarray}
The integrals appearing in eq.\,(\ref{eq:eq68}) can be computed as
\begin{eqnarray}
&& \int_0^{\infty} {\rm d}p \ p^2 f_{p \pm \bar \mu} \sin [2(\tau - \tau_{\rm i}) p] 
\nonumber \\
=&& \bar T^3 \int_0^{\infty} {\rm d}x \ x^2 \frac{\sin [2(\tau - \tau_{\rm i})\bar Tx]}{{\rm e}^{x \pm \bar \mu/\bar T} + 1}
\nonumber \\
\label{eq:eq69}
= && -{\rm i} \bar T^3 {\rm e}^{\mp \bar \mu/\bar T} \Bigl[ \Phi^{\ast}\bigl(-{\rm e}^{\mp \bar \mu/\bar T}, 3, 1 - 2{\rm i}(\tau - \tau_{\rm i}) \bar T\bigr) 
-\Phi^{\ast}\bigl(-{\rm e}^{\mp \bar \mu/\bar T}, 3, 1 + 2{\rm i}(\tau - \tau_{\rm i}) \bar T\bigr)  \Bigr] \nonumber \\
= && 2 \bar T^3 {\rm e}^{\mp \bar \mu/\bar T} \mathrm{Im} \,\Phi^{\ast}\bigl(-{\rm e}^{\mp \bar \mu/\bar T}, 3, 1 - 2{\rm i}(\tau - \tau_{\rm i}) \bar T\bigr), 
\label{eq:eq70}
\end{eqnarray}
where $\Phi^{\ast}(z, s, a)$ is the Lerch transcendent function, defined in eq.\,(\ref{eq:eq71}), and
\begin{gather}
\int_0^{\infty} {\rm d}p \ p (f_{p - \bar \mu} + f_{p + \bar \mu})  = - \bar T^2 \Bigl[\mathrm{Li}_{2} \bigl(-{\rm e}^{\bar \mu/\bar T}\bigr) + \mathrm{Li}_{2} \bigl(-{\rm e}^{-\bar \mu/\bar T}\bigr) \Bigr]\,,  
\label{eq:eq73}
\\
\int_0^{\infty} {\rm d}p \ p^2 f_{p + \bar \mu} (1 + f_{p + \bar \mu}) = -2 \bar T^3 \mathrm{Li}_{2} \bigl(-{\rm e}^{-\bar \mu/\bar T}\bigr)\,,
\label{eq:eq74}
\end{gather}
leading to the term linear in $k$ in eq.\,(\ref{eq:eq67}).

\section{Estimate of $\tau \bar{T}$ and $\tau_{\rm i} \bar{T}$}
\label{sec:estimate_tau}
We here derive an order-of-magnitude estimate of $\tau \bar T$ and $\tau_{\rm i} \bar T$ in the radiation dominated period, 
in which the scale factor evolves as (see, e.g., ref.~\cite{Yagi:2005yb})
\begin{eqnarray}
a(t) = A t^{1/2}, 
\label{eq:eq53}
\end{eqnarray}
where
\begin{eqnarray}
A = \left[ \frac{32 \pi G}{3} (\epsilon a^4) \right]^{1/4},  
\label{eq:eq54}
\end{eqnarray}
with $\epsilon a^4 =$ constant. 
Here, $\epsilon$ is the energy density, which is expressed at leading order of the $\mu/T$ expansion as
\begin{eqnarray}
\epsilon = g_{\ast}(T) \frac{\pi^2}{30} T^4\,,
\label{eq:eq60}
\end{eqnarray} 
where $g_{\ast}(T)$ is the number of relativistic degrees of freedom at temperature $T$. 

From the equations above, $t$ can be related to $T$ as 
\begin{eqnarray}
t = \left( \frac{45}{16 \pi^3 G g_{\ast}(T)} \right)^{1/2} \frac{1}{T^2}\,.
\label{eq:eq56}
\end{eqnarray}
Furthermore, we can obtain $\tau$ from ${\rm d} t = a {\rm d}\tau$ and eq.~(\ref{eq:eq53}) as 
\begin{eqnarray}
\tau-\tau_{\rm i} = \int^t_{t_{\rm i}} \frac{{\rm d} t}{a(t)} = \frac{2}{A} (t^{1/2}-t_{\rm i}^{1/2})\,,
\label{eq:eq55}
\end{eqnarray} 
and hence,
\begin{eqnarray}
\tau \bar T-\tau_{\rm i} \bar T &=& 2 \left( \frac{2025}{256 \pi^6 G^2 g^2_{\ast}(T) }  \right)^{1/4} \frac{1}{T}\left(1-\frac{\sqrt{g_{\ast}(T)}T}{\sqrt{g_*(T_{\rm i})}T_{\rm i}} \right). \label{eq:eq61.1} 
\end{eqnarray} 
By extracting the $\tau$-dependent component above, we find
\begin{eqnarray}
\tau \bar T=2 \left( \frac{2025}{256 \pi^6 G^2 g^2_{\ast}(T) }  \right)^{1/4} \frac{1}{T} \sim 10^{20} \left(\frac{T}{\mathrm{GeV}}\right)^{-1}.
\label{eq:eq61}
\end{eqnarray}
We can therefore conclude that $\tau \bar T \gg 1$, for relevant temperatures in the radiation dominant period. 

Considering next $\tau_{\rm i} \bar T$, 
based on the $\tau$-independent part in eq.~(\ref{eq:eq61.1}), 
we can derive 
\begin{eqnarray}
\label{eq:tau_i_bar_T_relation_2}
\tau_{\rm i} {\bar T} &=& 2 \left[ \frac{2025}{256 \pi^6 G^2 g^2_{\ast}(\tau_{\rm i})}  \right]^{1/4}  \frac{1}{T(\tau_{\rm i})}, 
\end{eqnarray}
which can further be simplified as 
\begin{eqnarray}
\label{eq:tau_i_bar_T_relation_3}
\tau_{\rm i} {\bar T} &=& \frac{3\sqrt{10}}{\pi g^{1/2}_{\ast}(\tau_{\rm i})}   \frac{m_{\mathrm{P}}}{T(\tau_{\rm i})}. 
\end{eqnarray}
The numerical factor in front of $m_{\mathrm{P}}/T(\tau_{\rm i})$ is approximately 1 and we hence have 
\begin{eqnarray}
\label{eq:tau_i_bar_T_relation_4}
\tau_{\rm i} {\bar T} &\simeq& \frac{m_{\mathrm{P}}}{T(\tau_{\rm i})}, 
\end{eqnarray}
which shows that $\tau_{\rm i} {\bar T} \gg 1$ as long as $T(\tau_{\rm i}) \ll m_{\rm P}$.

\end{document}